\documentclass[12pt,preprint]{aastex}
\usepackage{color}

\def\etal{~et al.}
\def\simlt{\lower.5ex\hbox{$\; \buildrel < \over \sim \;$}}
\def\simgt{\lower.5ex\hbox{$\; \buildrel > \over \sim \;$}}

\def\gsim{\lower 2pt \hbox{$\, \buildrel {\scriptstyle >}\over
{\scriptstyle \sim}\,$}}
\def\lsim{\lower 2pt \hbox{$\, \buildrel {\scriptstyle <}\over
{\scriptstyle \sim}\,$}}

\def\deg{\ifmmode ^{\circ}
         \else $^{\circ}$\fi}
\def\pdeg{\ifmmode
           $\setbox0=\hbox{$^{\circ}$}\rlap{\hskip.11\wd0 .}$^{\circ}
     \else \setbox0=\hbox{$^{\circ}$}\rlap{\hskip.11\wd0 .}$^{\circ}$\fi}
     
\def\pc{\ifmmode \mathrm{pc} \else $\mathrm{pc}$ \fi}
\def\mpc{\ifmmode \mathrm{Mpc} \else $\mathrm{Mpc}$\fi}
\def\mpcthree{\ifmmode \mathrm{Mpc}^{-3} \else $\mathrm{Mpc}^{-3}$\fi}
\def\gpcthree{\ifmmode \mathrm{Gpc}^{-3} \else $\mathrm{Gpc}^{-3}$\fi}

\def\kelvin{\ifmmode \mathrm{K} \else {$\mathrm{K}$}\fi}
\def\kev{\ifmmode \mathrm{keV} \else $\mathrm{keV}$ \fi}

\def\lsun{\ifmmode {L_\odot} \else $L_\odot$\fi}
\def\msun{\ifmmode M_\odot \else $M_\odot$\fi}
\def\msunyr{\ifmmode M_\odot~\mathrm{yr}^{-1} \else $M_\odot~\mathrm{yr}^{-1}$\fi}

\def\cosi{\ifmmode {\cos\,i} \else $\cos\,i$\fi}

\def\heii{\ifmmode {\rm He{\sc ii}} \else He~{\sc ii}\fi}
\def\mgii{\ifmmode {\rm Mg{\sc ii}} \else Mg~{\sc ii}\fi}
\def\caii{\ifmmode {\rm Ca{\sc ii}} \else Ca~{\sc ii}\fi}
\def\ciii{\ifmmode {\rm C{\sc iii}]} \else C~{\sc iii}]\fi}
\def\civ{\ifmmode {\rm C{\sc iv}} \else C~{\sc iv}\fi}
\def\mgii{\ifmmode {\rm Mg{\sc ii}} \else Mg~{\sc ii}\fi}
\newcommand{\oi}{{\sc [O~i]}}
\newcommand{\oii}{{\sc [O~ii]}}
\newcommand{\oiii}{{\sc [O~iii]}}

\newcommand{\neiii}{{[Ne~{\sc iii}]}}
\newcommand{\nev}{{[Ne~{\sc v}]}}
\newcommand{\sii}{{\sc [S~ii]}}

\def\teff{\ifmmode {T_{\rm eff}} \else $T_{\rm eff}$\fi}
\def\tmax{\ifmmode {T_{\rm max}} \else $T_{\rm max}$\fi}

\def\mbh{\ifmmode {M_{\rm BH}} \else $M_{\rm BH}$\fi}
\def\led{\ifmmode L_{\mathrm{Ed}} \else $L_{\mathrm{Ed}}$\fi}
\def\lbolflare{\ifmmode L_{\mathrm{bol,flare}} \else $L_{\mathrm{bol,flare}}$\fi}
\def\lagn{\ifmmode L_{\mathrm{agn}} \else $L_{\mathrm{agn}}$\fi}
\def\lbolagn{\ifmmode L_{\mathrm{bol,agn}} \else $L_{\mathrm{bol,agn}}$\fi}
\def\lbol{\ifmmode L_{\mathrm{bol}} \else $L_{\mathrm{bol}}$\fi}
\def\mdot{\ifmmode {\dot M} \else $\dot M$\fi}
\def\mdoto{\ifmmode {\dot{M}_0} \else  $\dot{M}_0$\fi}
\def\mdotf{\ifmmode {\dot{M}_\mathrm{flare}} \else  $\dot{M}_\mathrm{flare}$\fi}

\def\hnot{\ifmmode H_0 \else H$_0$ \fi}

\def\vkep{\ifmmode v_\mathrm{Kep} \else $v_\mathrm{Kep}$ \fi}
\def\vc{\ifmmode v_\mathrm{c} \else $v_\mathrm{c}$ \fi}

\def\vthree{\ifmmode v_{1000} \else $v_{1000}$ \fi}
\def\vrel{\ifmmode v_\mathrm{rel} \else $v_\mathrm{rel}$ \fi}
\def\vkick{\ifmmode v_\mathrm{kick} \else $v_\mathrm{kick}$ \fi}
\def\vkickz{\ifmmode v_{\mathrm{kick},z} \else $v_{\mathrm{kick},z} $ \fi}
\def\vkicky{\ifmmode v_{\mathrm{kick},y} \else $v_{\mathrm{kick},y} $ \fi}
\def\vchar{\ifmmode v_\mathrm{char} \else $v_\mathrm{char}$ \fi}
\def\eflare{\ifmmode E_\mathrm{flare} \else $E_\mathrm{flare}$ \fi}
\def\ekick{\ifmmode E_\mathrm{kick} \else $E_\mathrm{kick}$ \fi}
\def\ecoll{\ifmmode E_\mathrm{coll} \else $E_\mathrm{coll}$ \fi}
\def\ezero{\ifmmode E_\mathrm{0} \else $E_\mathrm{0}$ \fi}
\def\efac{\ifmmode \xi_\mathrm{E} \else $\xi_\mathrm{E}$ \fi}
\def\tqso{\ifmmode t_\mathrm{QSO} \else $t_\mathrm{QSO}$ \fi}
\def\tflare{\ifmmode t_\mathrm{flare} \else $t_\mathrm{flare}$ \fi}
\def\tzero{\ifmmode t_\mathrm{0} \else $t_\mathrm{0}$ \fi}
\def\tfac{\ifmmode \xi_\mathrm{t} \else $\xi_\mathrm{t}$ \fi}
\def\gfac{\ifmmode f_\mathrm{g} \else $f_\mathrm{g}$ \fi}
\def\lflare{\ifmmode L_\mathrm{flare} \else $L_\mathrm{flare}$ \fi}
\def\fflare{\ifmmode F_\mathrm{flare} \else $F_\mathrm{flare}$ \fi}
\def\nflare{\ifmmode N_\mathrm{flare} \else $N_\mathrm{flare}$ \fi}
\def\tshock{\ifmmode T_\mathrm{shock} \else $T_\mathrm{shock}$ \fi}
\def\rmin{\ifmmode R_\mathrm{1} \else $R_\mathrm{1}$ \fi}
\def\rmax{\ifmmode R_\mathrm{2} \else $R_\mathrm{2}$ \fi}
\def\rbound{\ifmmode R_\mathrm{b} \else $R_\mathrm{b}$ \fi}
\def\pbound{\ifmmode P_\mathrm{b} \else $P_\mathrm{b}$ \fi}
\def\mbound{\ifmmode M_\mathrm{b} \else $M_\mathrm{b}$ \fi}
\def\mbo{\ifmmode M_{\mathrm{b}0} \else $M_{\mathrm{b}0} $ \fi}
\def\ebo{\ifmmode E_{\mathrm{b}0} \else $E_{\mathrm{b}0} $ \fi}
\def\efinal{\ifmmode E_\mathrm{final} \else $E_\mathrm{final} $ \fi}
\def\tbound{\ifmmode t_\mathrm{b} \else $t_\mathrm{b}$ \fi}
\def\tagn{\ifmmode t_\mathrm{AGN} \else $t_\mathrm{AGN}$ \fi}
\def\torb{\ifmmode t_\mathrm{orb} \else $t_\mathrm{orb}$ \fi}
\def\tdf{\ifmmode t_\mathrm{df} \else $t_\mathrm{df}$ \fi}
\def\rlim{\ifmmode R_\mathrm{lim} \else $R_\mathrm{lim}$ \fi}
\def\vlim{\ifmmode v_\mathrm{lim} \else $v_\mathrm{lim}$ \fi}
\def\vphi{\ifmmode v_\phi \else $v_\phi$ \fi}
\def\mlim{\ifmmode M_\mathrm{lim} \else $M_\mathrm{lim}$ \fi}
\def\tlim{\ifmmode t_\mathrm{lim} \else $t_\mathrm{lim}$ \fi}
\def\llim{\ifmmode L_\mathrm{lim} \else $L_\mathrm{lim}$ \fi}
\def\fqso{\ifmmode f_\mathrm{QSO} \else $f_\mathrm{QSO}$ \fi}

\def\hbeta{\ifmmode \rm{H}\beta \else H$\beta$\fi}
\def\hbetan{\ifmmode \rm{H}\beta_{\rm n} \else H$\beta_{\rm n}$\fi}
\def\hgamma{\ifmmode \rm{H}\gamma \else H$\gamma$\fi}
\def\hdelta{\ifmmode \rm{H}\delta \else H$\delta$\fi}
\def\hepsilon{\ifmmode \rm{H}\epsilon \else H$\epsilon$\fi}
\def\hzeta{\ifmmode \rm{H}\zeta \else H$\zeta$\fi}
\def\halpha{\ifmmode \rm{H}\alpha \else H$\alpha$\fi}
\def\lalpha{\ifmmode \rm{Ly}\alpha \else Ly$\alpha$}

\def\dvhb{\ifmmode \Delta v_{\hbeta} \else $\Delta v_{\hbeta}$\fi}
\def\dvmg{\ifmmode \Delta v_{\rm{Mg}} \else $\Delta v_{\rm{Mg}}$\fi}

\def\muobs{\ifmmode {\mu_{o}} \else  $\mu_{o}$ \fi}
\def\cosi{\ifmmode {\mathrm{cos}\,i} \else $\mathrm{cos}\,i$\fi}

\def\teff{\ifmmode {T_{eff}} \else $T_{eff}$ \fi}
\def\tmax{\ifmmode {T_{max}} \else $T_{max}$ \fi}

\def\tauh{\ifmmode {\tau_{\rm H}} \else $\tau_{\rm H}$ \fi}

\def\yr{\ifmmode {\rm yr} \else  yr \fi}
\def\kms{\ifmmode \rm km~s^{-1}\else $\rm km~s^{-1}$\fi}
\def\cm{\ifmmode {\rm cm} \else  cm \fi}
\def\cmmitwo{\ifmmode \rm cm^{-2} \else $\rm cm^{-2}$\fi}
\def\cmmithree{\ifmmode \rm cm^{-3} \else $\rm cm^{-3}$\fi}
\def\cmps{\ifmmode \rm cm~s^{-1}\else $\rm cm~s^{-1}$\fi}
\def\cmpsps{\ifmmode \rm cm~s^{-2}\else $\rm cm~s^{-2}$\fi}
\def\kmps{\ifmmode \rm km~s^{-1}\else $\rm km~s^{-1}$\fi}
\def\kmpspmpc{\ifmmode \rm km~s^{-1}~Mpc^{-1} \else
    $\rm km~s^{-1}~Mpc^{-1}$\fi}
  
\def\gcmthree{\ifmmode \rm g~cm^{-3} \else $\rm g~cm^{-3}$\fi}
\def\gcmtwo{\ifmmode \rm g~cm^{-2} \else $\rm g~cm^{-2}$\fi}
   
\def\erg{\ifmmode {\rm erg} \else $\rm erg$ \fi}
\def\ergps{\ifmmode {\rm erg~s^{-1}} \else $\rm erg~s^{-1}$ \fi}
\def\ergcms{\ifmmode \rm erg~cm^{-2}~s^{-1} \else $\rm erg~cm^{-2}~s^{-1}$ \fi}
\def\ergcmshz{\ifmmode \rm erg~s^{-1}~cm^{-2}~Hz^{-1} \else $\rm
erg~cm^{-2}~s^{-1}~Hz^{-1}$ \fi}
\def\ergcmsa{\ifmmode \rm erg~cm^{-2}~s^{-1}~\AA^{-1} \else $\rm
erg~cm^{-2}~s^{-1}~\AA^{-1}$ \fi}
\def\ergshz{\ifmmode \rm erg s^{-1} Hz^{-1} \else
   $\rm erg s^{-1} Hz^{-1}$ \fi}

\def\lam{\ifmmode {\lambda} \else {$\lambda$} \fi}
\def\llam{\ifmmode {L_\lambda} \else  $L_\lambda$ \fi}
\def\lamLlam{\ifmmode \lambda L_{\lambda}(5100) \else {$\lambda L_{\lambda}(5100)$} \fi}
\def\nuLnu{\ifmmode \nu L_{\nu}(5100) \else {$\nu L_{\nu}(5100)$} \fi}
\def\ilam{\ifmmode {I_\lambda} \else  $I_\lambda$ \fi}
\def\flam{\ifmmode {F_\lambda} \else  $F_\lambda$ \fi}
\def\inu{\ifmmode {I_\nu} \else  $I_\nu$ \fi}
\def\fnu{\ifmmode {F_\nu} \else  $F_\nu$ \fi}
\def\bnu{\ifmmode {B_\nu} \else  $B_\nu$ \fi}

\def\msigma{\ifmmode M_{\sigma} \else $M_{\sigma}$\fi}
\def\mbulge{\ifmmode M_{\mathrm{bulge}} \else $M_{\mathrm{bulge}}$\fi}
\def\mgal{\ifmmode M_{\mathrm{gal}} \else $M_{\mathrm{gal}}$\fi}
\def\lgal{\ifmmode L_{\mathrm{gal}} \else $L_{\mathrm{gal}}$\fi}
\def\lbulge{\ifmmode L_{\mathrm{bulge}} \else $L_{\mathrm{bulge}}$\fi}
\def\mgalstar{\ifmmode M^*_{\mathrm{gal}} \else $M^*_{\mathrm{gal}}$\fi}

\def\mbhsigstar{\ifmmode M_{\mathrm{BH}} - \sigma_* \else $M_{\mathrm{BH}} - \sigma_*$ \fi}
\def\deltalogmbh{\ifmmode \Delta~{\mathrm{log}}~M_{\mathrm{BH}} \else $\Delta$~log~$M_{\mathrm{BH}}$\fi}

\def\sigstar{\ifmmode \sigma_* \else $\sigma_*$\fi}
\def\sigthree{\ifmmode \sigma_{\mathrm{[O~III]}} \else $\sigma_{\mathrm{[O~III]}}$\fi}
\def\sigtwo{\ifmmode \sigma_{\mathrm{[O~II]}} \else $\sigma_{\mathrm{[O~II]}}$\fi}
\def\signl{\ifmmode \sigma_{\mathrm{NL}} \else $\sigma_{\mathrm{NL}}$\fi}
\def\wthree{\ifmmode {\rm FWHM({[O~III]})} \else $FWHM({[O~III]})$ \fi}
\def\wtwo{\ifmmode {\rm FWHM({[O~II]})} \else $FWHM({[O~II]})$ \fi}
\def\mthree{\ifmmode M_{\mathrm [O~III]} \else $M_{\mathrm [O~III]}$ \fi}
\def\mtwo{\ifmmode M_{\mathrm [O II]} \else $M_{\mathrm [O II]}$ \fi}

\def\lbreak{\ifmmode L_{\mathrm{break}} \else $L_{\mathrm{break}}$\fi}
\def\lcut{\ifmmode L_{\mathrm{cut}} \else $L_{\mathrm{cut}}$\fi}

\received{}
\accepted{}
\journalid{}{}
\articleid{}{}

\slugcomment{Submitted to ApJ}

\shortauthors{Smith, Shields, Bonning, McMullen, Rosario \& Salviander}
\shorttitle{Binary QSOs}

\begin{document}

\title{A Search for Binary Active Galactic Nuclei: Double-Peaked \oiii\ AGN in the Sloan Digital Sky Survey}

\author{K.~L. Smith\altaffilmark{1}, G.~A. Shields\altaffilmark{1}, E.~W. Bonning\altaffilmark{2}, C.~C. McMullen\altaffilmark{1}, D.~J. Rosario\altaffilmark{3}, S.~Salviander\altaffilmark{1}}

\altaffiltext{1}{Department of Astronomy, University of Texas, Austin,
TX 78712; krista@mail.utexas.edu, shields@astro.as.utexas.edu, triples@astro.as.utexas.edu} 

\altaffiltext{2}{YCAA - Department of Physics, Yale University, New Haven, CT 06520; erin.bonning@yale.edu}

\altaffiltext{3}{Lick Observatory, University of California, Santa Cruz 95064; rosario@ucolick.org}

\begin{abstract}

We present AGN from the Sloan Digital Sky Survey (SDSS)  having double-peaked profiles of \oiii~$\lambda\lambda~5007,4959$ and other narrow emission-lines, motivated by the prospect of finding candidate binary AGN. These objects were identified by means of a visual examination of 21,592 quasars at $z<0.7$ in SDSS Data Release 7 (DR7). Of the spectra with adequate signal-to-noise, 148 spectra exhibit a double-peaked \oiii\ profile. Of these, 86 are Type 1 AGN and 62 are Type 2 AGN. Only two give the appearance of possibly being optically resolved double AGN in the SDSS images, but many show close companions or signs of recent interaction.
Radio-detected quasars are three times more likely to exhibit a double-peaked \oiii\ profile than quasars with no detected radio flux, suggesting a role for jet interactions in producing the double-peaked profiles.   Of the 66 broad line (Type 1) AGN that are undetected in the FIRST survey,
0.9\% show double peaked \oiii\ profiles.   We discuss statistical tests of the nature of the double-peaked objects.  Further study is needed to determine which of them are binary AGN rather than disturbed narrow line regions, and how many additional binaries may remain undetected because of insufficient line-of-sight velocity splitting.

Previous studies indicate that 0.1\%\ of SDSS quasars are spatially resolved binaries,
with typical spacings of $\sim10$  to 100~kpc.  If a substantial fraction of the double-peaked objects are indeed binaries, then our results imply that binaries occur more frequently at smaller separations ($< 10$~kpc).  This suggests that simultaneous fueling of both black holes is more common as the binary orbit decays through these spacings.

\end{abstract}


\section{Introduction}
\label{sec:intro}

Binary quasars are a rare but important aspect of galactic evolution and the
AGN phenomenon, as reviewed by \citet{komossa06}.  Studies of the
nature and incidence of binary quasars provide insight into galaxy
mergers at substantial look-back times and into the fueling of AGN
by accretion onto supermassive black holes. According to current cosmological models, 
many large galaxies have undergone at least one major merger. 
A likely mechanism for AGN fueling is the migration of gas caused by tidal torques in a merger.  
Therefore, one might expect to see a substantial number of binary quasars.

Evidence for binary AGN includes peculiar morphologies of radio galaxies,
X-ray resolved double nuclei, and optically resolved double QSOs.
Out of $10^{5} $ QSOs, only $\sim10^2$ optical binary quasars are known
\citep[and references therein]{foreman09}.
Most have spacings of 10~kpc or more, and lie at $z\sim2$ to $z\sim3$, where study of
the host galaxy is difficult. Binaries with smaller separations  include a 2~kpc (0.3~arcsec) optical example at $z = 0.848$ observed with HST by \citet{junkkarinen01}, and the ULIRG NGC 6240 with a double X-ray nucleus separated by 1.4~kpc \citep{komossa03}.  Both authors argue  that the observed incidence of such close binaries is roughly 100 times less than might be expected if the fueling of one black hole giving QSO activity at the relevant stages of a merger mandates the fueling of the second black hole.  A single, much closer (7~pc) binary AGN candidate is the radio galaxy 0402+379  \citep{rodriguez09}. Additional examples of optically resolved binary AGN include a z=0.36 galaxy from the COSMOS survey with a spacing of 2.5 kpc \citep{comerford09b} and a z=0.44 binary quasar with a spacing of 21 kpc \citep{green10}. 

The narrow emission lines of AGN offer a potential indicator of binary AGN at intermediate separations of $\sim 1$ to 10~kpc.  The radius of the narrow emission-line region (NLR) is typically a few hundred parsecs, so that such objects should have distinct NLRs orbiting with each of the merging nuclei with its black hole.
Typical orbital velocities in galaxies are of order a few hundred kilometers per second, similar to the widths  of AGN narrow lines.  If two adjacent narrow-line regions in a merging system were detected by a single spectrograph slit or fiber, a double-peaked emission-line profile could result.   This provides a technique for identifying candidate binary AGN that are spatially unresolved.  Several examples have already been discussed in the literature \citep{zhou04, gerke07}.    \citet{comerford09a} have studied AGN with narrow lines displaced in
velocity from the host galaxy as possible examples of galactic mergers in progress, including two objects with double-peaked narrow lines.

We have carried out a search for double-peaked narrow line profiles, primarily the \oiii\ lines, in quasar spectra from the Sloan Digital Sky Survey\footnote{The SDSS website is
http://www.sdss.org.}.  The goals of this work were to identify interesting individual objects for study, and to constrain the frequency of binary quasars in the relevant regime of separation.  This search has identified 148 double-peaked \oiii\ objects, with velocity splittings between 180~\kms\ and 1400~\kms. This includes roughly equal numbers of AGN 1 (broad lines) and AGN 2 (narrow lines only).  Here we describe the search and the double-peaked objects, consider alternative causes of the double-peaked profiles, including bipolar jets and disk rotation, and discuss statistical inferences for AGN fueling in merging galactic nuclei.
We assume a concordance cosmology with H$_{0}=70$ \kms Mpc$^{-1}$, 
$\Omega_{\Lambda}=0.7$.

Recently, two independent studies by \citet{liu09} and \citet{wang09} have appeared, involving narrow line AGN 2 selected from the SDSS galaxy database.
These authors find a similar incidence of double \oiii\ objects to that found here.  The two studies are
complementary to ours, in that our work emphasizes broad line objects.

\section{Sample and Method}
\label{sec:sample}

Our spectra are obtained from SDSS DR7, which contains 21,592 QSO spectra within our redshift range, $0.1 < z < 0.7$. Because it is possible to have strong \oiii\ emission despite a noisy spectrum, we imposed no {\em a priori}   signal-to-noise (S/N) cutoff.  We conducted a visual inspection of the spectra in the region of several key emission lines. Our primary search criterion was a double-peaked \oiii\ profile in both $\lambda 5007$ and $\lambda 4959$,  consistent with the 3:1 intensity ratio fixed by atomic physics.  Once these objects had been identified, a more thorough inspection of each double \oiii\ object was conducted with the IRAF routine SPLOT \footnote{IRAF is distributed by the National Optical Astronomy Observatories, which are operated by the Association of Universities for Research in Astronomy, Inc., under cooperative agreement with the National Science Foundation.}  to determine whether the velocity splitting observed in \oiii\ was detectable in other emission lines. The lines considered were \hbeta\ , \oiii\ , \oii\ , \neiii\ , and \sii. While confirmation of the \oiii\ split in these lines strengthens the case, we did not require it for inclusion in the list of double-peaked objects if the S/N was such that the double-peak was not ruled out. For approximately 12 of our objects, the \nev\ line showed sufficient S/N to allow measurement of the split.  In these cases, there was agreement with the \oiii\ velocity split within about 10\%, similar to the measurement accuracy.  
In a few more cases, the S/N allowed qualitative confirmation of the split by visual inspection.  It appears
that the split is seen in \nev\ whenever the S/N is adequate to see it,  Thus, the double-peaked profile extends to higher ionization lines in most if not all cases.  However, \nev\ was rarely useful to confirm an \oiii\ split that was not already confirmed by other lines.

Objects were grouped into two categories, ``good" and ``marginal." Marginal objects are questionable cases due to low S/N, bad sky subtraction, and most often, failure of $\lambda 4959$ to reproduce the intensity ratio of the components in $\lambda 5007$. Our objects are listed in \ref{t:tab1} and their radio properties (see discussion below) are given in \ref{t:tab2}.

\begin{figure}[ht]
\begin{center}
\plotone{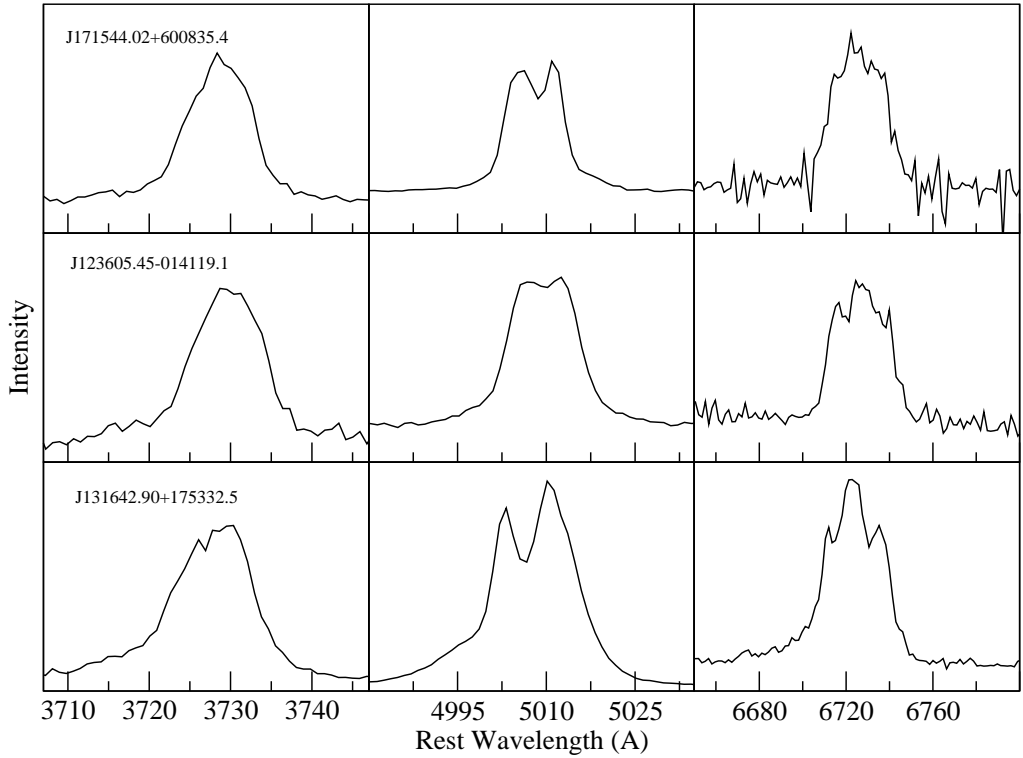}
\figcaption[]{The distinctive 1-2-3 pattern typical of many of our objects. The leftmost panels show the \oii\ lines, nearly always single-peaked. The center panels display the double-peaked \oiii\ lines. The last panels show the triple-peaked structure of the \sii\ line, discussed in the text. 
\label{fig:sulfur} }
\end{center}
\end{figure}

Generally, the double-peaked profile was confirmed in lines such as \neiii~$\lambda3869$ and the narrow \hbeta\ component when the S/N was adequate.  For the \sii~$\lambda\lambda6716, 6730$, the typical  \oiii\ double peak splitting of $\sim400$ or $500~\kms$ is such that the $\lambda6730$ of the lower redshift component falls on top of the $\lambda6716$ line of the higher redshift component.  This leads to a distinctive triple-peaked structure that was often seen when \sii\  was within the spectral range.  For \oii~$\lambda\lambda3726, 3729$, the profile rarely showed a double peak.  The separation of the $\lambda3726$ and $\lambda3729$ lines is $225~\kms$.  For a typical \oiii\ velocity split of $\sim400~\kms$, the $\lambda3726$ line of the higher redshift component fills in the gap between the  $\lambda3726$ and $\lambda3729$ lines of the lower redshift component.   This often gives a single peaked profile.  To verify this, we computed a simulated \oii\ blend for a typical candidate quasar using the relative intensity and velocity separation of the double-peaked \oiii\ profile.  The simulated \oii\ profile does not show the double peak.  (In a few cases, the splitting is seen in \oii.  The most dramatic case is in J0126+14, where the \oii\ is clearly resolved into two individual lines with the same 580~\kms\ velocity splitting as in \oiii). Because \oii\ is almost never resolved into two components, \oiii\ is split in all of our objects, and \sii\ often exhibits a triple-peaked structure, a distinctive 1-2-3 pattern is often seen in our objects if \sii\ is within the spectral range: one peak on \oii, two peaks on \oiii, and three peaks on \sii, as seen in Figure \ref{fig:sulfur}.

To confirm that both velocity components are typically present in \oii, we measured the centroid wavelength of the \oii\ feature and compared this redshift with the redshift of the blue and red components of the double-peaked \oiii\ profile.  For a control sample of several normal QSOs with single-peaked \oiii, the \oii\ and \oiii\ redshifts agreed to within $\sim15~\kms$, assuming a
typical doublet ratio $I(\lambda3726)/I(\lambda3729) = 0.83$ \citep{salviander07}.   Very few of our double-peaked \oiii\ objects were consistent with an \oii\ doublet being present solely at either the red or blue \oiii\ redshift.  For the double peaked \oiii\ objects, the \oii\ centroid was uniformly distributed between the red and blue \oiii\ redshifts, consistent with the fact that the primary (stronger) \oiii\ peak can be either the red or blue peak with similar frequency.  The \oii\ centroid redshift was on average 37\% of the way from the primary \oiii\ redshift to the secondary \oiii\ redshift, implying that the stronger \oiii\ component corresponds to the
stronger \oii\ component in most objects.   \citet{zhou04} note the absence of a double peak for \oii\ in most double-peaked \oiii\ AGN, and suggest that this can be reconciled with a binary nature for these objects if the \oii\ region is spatially more extended than for \oiii.  Our results suggest that \oii\ does typically show both \oiii\ velocity components, and no such explanation is necessary.

Many of our objects, including a number of the AGN 1, show stellar absorption lines from the host galaxy in their spectrum.  This affords an opportunity to examine the velocity of the \oiii\ components relative to the velocity of the host, as done by \citet{liu09} and \citet{wang09} for their AGN 2 with double-peaked \oiii. In particular, Wang et al. offer a possible ``virial'' test of binarity involving the relative fluxes and velocities of the \oiii\ components.  In order to estimate fluxes of the \oiii\ components, and to
have a measure of the velocities independent of the cursor measurements described above, we carried out double Gaussian fits to the spectra of our double-peaked objects.  The procedure assumed two Gaussians for the two components of the $\lambda5007$ and
$\lambda4959$ lines, with adjustable velocity, FWHM, and flux, and the 3:1 intensity ratio required by atomic physics.  The continuum in the vicinity of \oiii\ was fit with a second-order polynomial.  
The fit was optimized in a least-squares fashion.
This procedure achieved a good fit to the total observed line profile in 106 of the 148 objects.   For some of these objects, however, a visual inspection suggests that there could be substantially different choices for the amount of flux in the two components.   For 85 of the 106 successful fits,  we judged that the allocation of flux to the two components was unambiguous; these values are given in in Table \ref{t:tab3}.   We found reasonable agreement between the
velocities measured with SPLOT and the Gaussian fits.  For example, the difference between the SPLOT and Gaussian redshifts of the red \oiii\ component  had a mean of 16~\kms\ and an r.m.s. scatter of 50~\kms\ in the rest frame.  In most of the cases with a larger discrepancy, inspection of the fits suggested that the SPLOT measurements were more reliable, perhaps because the \oiii\ line is typically more sharply peaked than a Gaussian.  We have used the SPLOT velocities below; use of the Gaussian velocities would not alter our conclusions.

Host galaxy velocities were determined by cross-correlating a template galaxy spectrum with the observed spectrum 
with the aid of the IRAF routine FXCOR used in interactive mode.  
All 148 of the double \oiii\ spectra were visually examined in the region containing the MgB and NaD lines. If
these features were visually detected (73 objects), the spectrum was tagged and converted to a linear
wavelength scale spanning 3600-8500~\AA\ with dispersion of 2.0 \AA/pixel.
The cross-correlation template was generated with the same wavelength scale and 
dispersion, using an early type galaxy SDSS spectral template.
FXCOR was used to cross-correlate each galaxy spectrum with the template. 
A cubic spline was used to model and subtract out the mean continuum shape and a
cosine bell was used to apodize 1\% of each 1D spectrum at both ends. Interactive
options were used to cross-correlate the spectra only around the MgB-NaD region
of the spectrum. If a well defined cross-correlation peak was not visible, the
cross-correlation output was rejected.
This procedure was successful for 67 objects, of which 46 have \oiii\ component fluxes in Table \ref{t:tab3}. 
Table \ref{t:tab1} gives the host galaxy redshift. The host redshifts have an average uncertainty of 45~\kms\ in the rest frame,
with a maximum uncertainty of 95~\kms.

\section{Double-Peaked \oiii\ Objects}
\label{sec:cand}

\begin{figure}[ht]
\begin{center}
\plotone{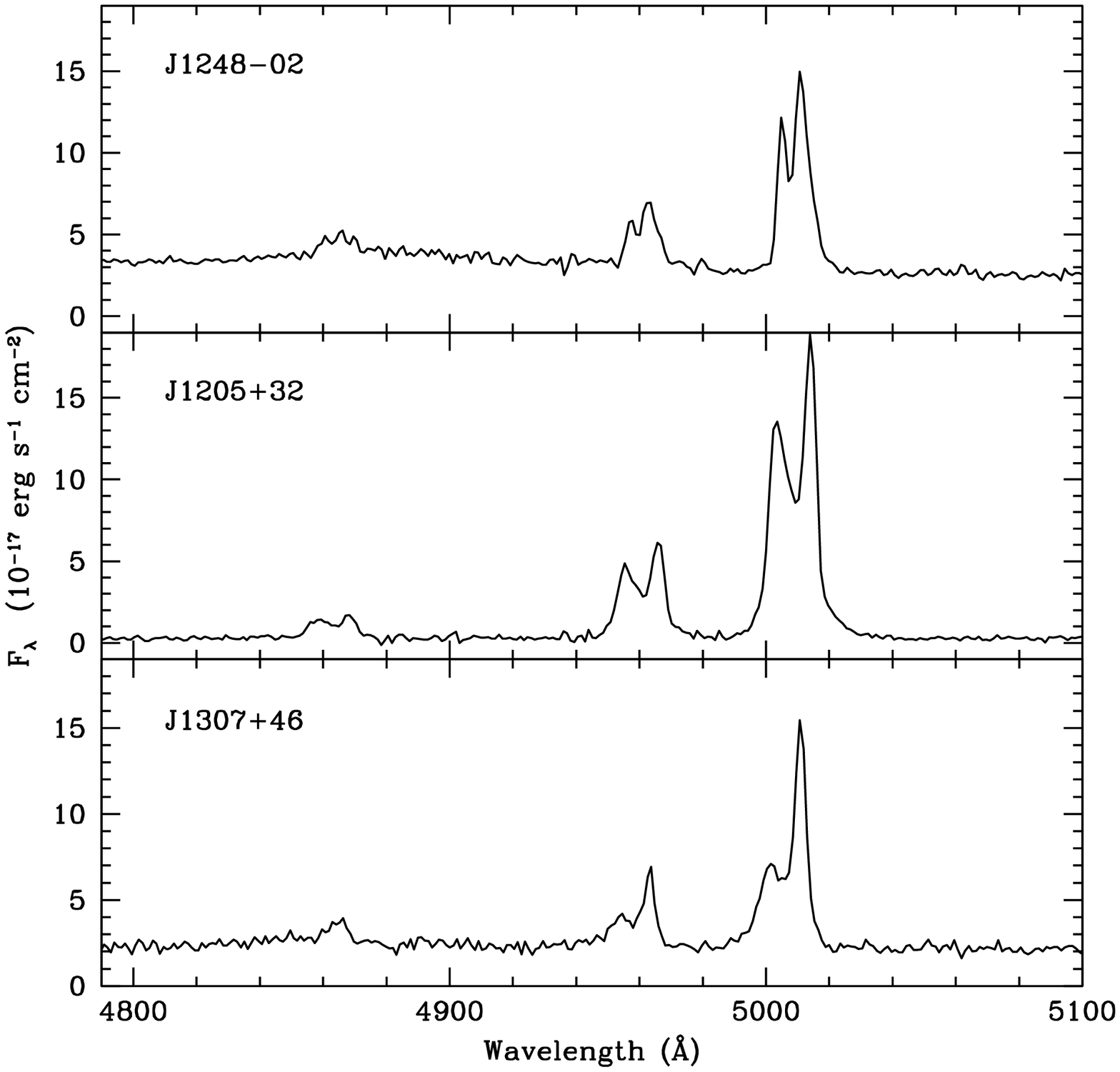}
\figcaption[]{
Spectra in the \oiii\ and \hbeta\ region for  three examples of objects with double-peaked \oiii\ 
profiles from Table 1.  Figure 3 gives the SDSS image of J1307+46.   In our full sample, the red and blue peaks are stronger with roughly equal frequency.
See text for discussion.
\label{fig:spectra} }
\end{center}
\end{figure}

We have a total of 148 double-peaked \oiii\ objects, listed in Table \ref{t:tab1}.   Objects with an asterisk exhibit either binarity or signs of interaction in the Sloan image.  Also given are the SDSS redshift, the velocity separation of the \oiii\ peaks, the continuum luminosity \nuLnu, and the AGN type (1 if broad emission lines are visible).  The ``Quality" column denotes whether the object is good, $g$, or marginal, $m$. There are 78 good and 70 marginal objects. While marginal objects do have a listed velocity splitting, these measurements are uncertain in some cases. Figure \ref{fig:spectra} gives spectra for three representative objects.

\begin{figure}[ht]
\begin{center}
\plotone{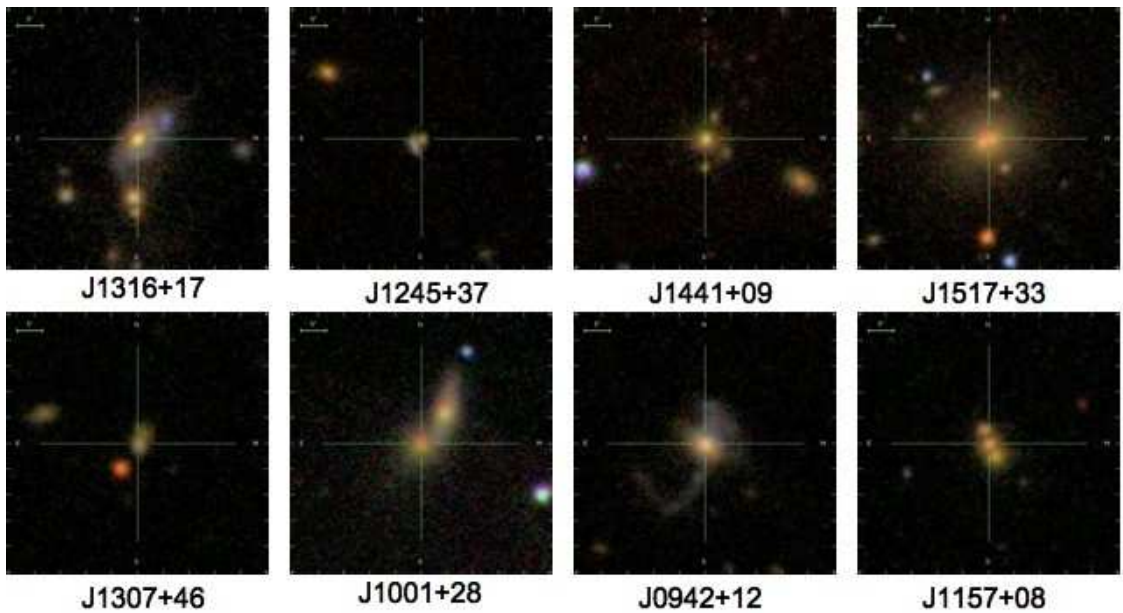}
\figcaption[]{
Eight SDSS images are displayed, corresponding to the eight objects in Table 1 with asterisks following their names. The scale bar in the upper left of the pictures corresponds to 5 arcseconds. The images exhibit objects with two or more obvious components, double nuclei, or signs of recent interaction. See text for discussion.
\label{fig:images} }
\end{center}
\end{figure}

Figure \ref{fig:images} gives the SDSS multi-color images of the eight visually interesting objects.
Two of these, J1316+17 and J1441+09, are included because they exhibit numerous smaller companions within a close radius.  Three of the images appear to consist of two or more individual galaxies at very close proximity: J1245+37 and J1307+46 appear to show pairs of galaxies centered on the SDSS fiber, and J1157+08, shows three galaxies in a linear arrangement. 

J0942+12 exhibits large tidal tails in the image, suggesting recent interaction. However, the nucleus itself does not appear binary.

J1001+28 may be a major merger between a large spiral and a spiral or elliptical. While the fiber width is not enough to encompass the NLRs of both objects, the galaxy is clearly undergoing interaction.

J1516+18 (not shown) has 1.3 Jy of radio flux, an order of magnitude larger than any other object in our sample. The object is a good double \oiii, although the velocity split which is  clear in \oiii\ is not seen in other lines. 

J1517+33 is discussed at length in \citet{rosario10}. It has emission line knots that straddle the stellar nucleus in velocity and position.  The double peaked profile appears to be the result of a bipolar radio jet.

\section{Statistical Results}
\label{sec:stats}

We did not impose an {\em a priori} S/N cutoff, in order to avoid missing noisy but interesting objects with strong \oiii\ lines.  However, some of the SDSS spectra did not have sufficient S/N to show the typical \oiii\ double peaks displayed by our objects.
Therefore, for statistical analysis, we must estimate what fraction of the full quasar sample had sufficient quality to be considered the parent sample of our double-peaked objects.  We find by inspection of a substantial subset of the 21,592 objects that only $40\%\pm10\%$ of spectra are viable in the sense that the double peaks of $\lambda 4959$, as present in a typical ``good'' object, would be clearly distinguishable from the noise.  While subjective, this estimate is reasonably sound because of the strong contrast between the S/N of the better and worse spectra.  Moreover, there are larger uncertainties in the analysis below, such as corrections for binaries with line-of-sight velocities too small to give a double-peaked \oiii\ profile (see below).

An issue for statistical analysis is the presence of some AGN 2 in the SDSS quasar download, numbering about 2.1\% of the full sample. Interestingly, 42\%\ of our objects are narrow-line objects, so that the Type 2 objects are twenty times overrepresented among our double \oiii\ objects.  We believe this to be a selection effect.  SDSS defines a quasar as having one or more emission lines with FWHM of at least 1000~\kms, intended to eliminate narrow-line objects. Evidently, the pipeline failure responsible for the presence of some AGN 2 in the parent sample is influenced by a double-peaked profile. Because of this dramatic selection effect, we omit the Type 2 objects from our statistical analysis.

After subtracting 2.1\% for the Type 2 objects, we have a
parent sample of 8,452 viable Type 1 spectra.  Our 86 Type 1 double \oiii\ objects amount to 1.0\%\ of the parent sample;  42 of these (0.50\%\ of the parent sample) are ``good". Our incidence of 1.0\%\ double-peaked AGN~1 is similar to the incidence of 1.1\%\ for AGN~2 in the study of \citet{liu09}.

Our double-peaked AGN were cross-referenced with the FIRST radio catalogue \citep{becker95}.  Only four of our objects were outside the footprint of the FIRST survey. FIRST detects 20~cm radio flux for 27\%\ of our Type 1 objects (detection limit 1 mJy),  as compared to only 9\%\ FIRST detections in the overall SDSS quasar catalogue in our redshift range \citep{schneider07}. In other words, radio detection is three times overrepresented among our double-peaked \oiii\ objects.  This suggests that radio jet interaction is often the cause of the double-peaked \oiii\ profiles.  Such interactions in the NLR are well known \citep[e.g.,][and references therein]{whittle04, whittle05}.  One of our double \oiii\ objects, J1517+33, appears to involve such an interaction.  It is a narrow-line object with a FIRST radio flux of 107~mJy and two bright  optical regions near its nucleus in the Sloan image.  Subsequent observations of this object with the VLA have shown a double-lobed radio structure at the same orientation and position as the bright optical regions. This object is discussed in \citet{rosario10}. 

We give the radio fluxes, luminosities, and  the \citet{kellermann89} radio loudness parameter in Table \ref{t:tab2}. We adopt the radio-loudness conventions used in \citet{kellermann89}. The ratio $R=F_r/F_o$ measures the ratio of 6 cm radio flux density to $4400$~\AA\ optical flux density. If $0.1<R<1$, we call an object radio quiet, $1<R<10$ is radio intermediate, and $10<R<100$ is radio loud. If an object is not detected in the FIRST sample, we refer to it as radio undetected. It should be noted that the lack of a FIRST detection does not indicate a total absence of radio flux, or of radio jets. High redshift objects with no FIRST detection may have marginal flux which could have been detected at lower redshifts, and orientation of the radio source can affect detected flux.  Therefore some objects with radio jets may fall below the FIRST detection limit. 

Additionally, Type 2 spectra are dominated by the light of the host galaxy, which clouds the meaning of the radio-to-optical flux. Consequently, the $R$ ratios given in Table \ref{t:tab2} for Type 2 objects should be taken with caution. Because our study focuses on broad-line objects and we omit the AGN 2 from our statistical interpretation, this issue does not affect our conclusions.

Because of the likelihood of radio jets causing the double-peaked signature, we consider our best candidates for true binaries to be those with no FIRST radio detections. Discarding 9\%\  of the parent sample to allow for FIRST-detected objects, our parent sample consists of 7,692 quasars.  Our double-peaked sample consists of 66 radio undetected broad-line objects, those we consider to be the best prospects. This is 0.9\%\ of the total number of viable SDSS radio undetected broad-line quasars. If we use only the ``good" objects, 29 remain, which is 0.38\%\  of the parent sample.

\begin{figure}[ht]
\begin{center}
\plotone{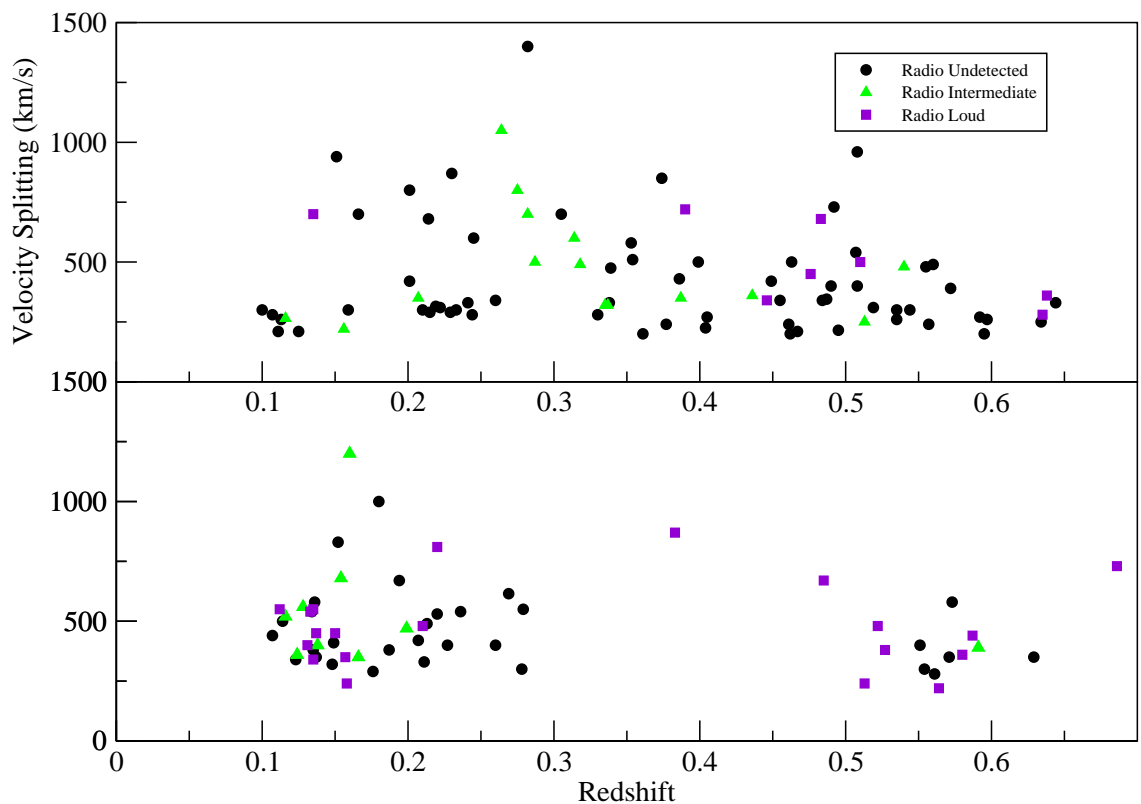}
\figcaption[]{
Velocity splitting of the type 1 (upper panel) and type 2  double-\oiii\ objects as a function of redshift.  There appears to be a slight trend toward fewer large splittings at larger redshifts. The graph is color-coded for radio-loudness, where this parameter is determined by the Kellerman $R$ ratio.  The gap at $z \sim 0.3$ to 0.5 for the AGN 2 is an artifact.  Note that radio detection by FIRST is redshift dependent.  See text for discussion.
\label{fig:redshift} }
\end{center}
\end{figure}

Figure \ref{fig:redshift} shows the relationship between the redshift and the velocity split for our double-peaked AGN. Objects at greater distance are typically at higher luminosity, and more luminous objects tend to have wider \oiii\ lines \citep[and references therein]{salviander07}. Accordingly, a larger velocity difference would be necessary for a resolved double-peaked profile. It would make sense, therefore, that as redshift increases, observed velocity splits should also increase. However, we find that the opposite is true. It appears that at high redshifts, the typical velocity splitting actually decreases, although the statistical sample is small.

When the redshift plot is made separately for AGN 1 and AGN 2, a distinctive gap is seen among the AGN 2 between redshifts 0.3 and 0.5. 
In a control sample of the SDSS AGN 2 in our quasar download, we also see a gap in that redshift range. We assume therefore that the gap results from the failure in the SDSS pipeline that allowed some AGN 2 into the quasar sample in the first place.

For our entire set of double-peaked objects, the redshifts range from $z = 0.100$ to 0.686 with a mean of 0.33.  For our ``good'' AGN~1, the average EW is 86~\AA.
 For the Type~1 objects with host galaxy redshifts,
 the midpoint in velocity between the red and blue components averages  $+33 \pm 29~\kms$ (standard error of mean), with an r.m.s. scatter of 108~\kms.  
 For our AGN~1 with \oiii\ component fits, the mean value of the Gaussian width parameter for the red component is $\sigma_{\mathrm r} = 143 \pm 15~\kms$,
 with an r.m.s. scatter of 57~\kms; for the blue component,  the mean is $\sigma_{\mathrm b} = 161\pm 18~\kms$, with a scatter of 66~\kms.  For a Gaussian,
 the FWHM is $2.35\sigma$.  For this group of objects, the mean separation is $\Delta V = 364~\kms$ which is 1.02 times the mean FWHM of either component.
 For comparison, \citet{liu09} found a typical $\Delta V$ larger by a factor $\sim1.5$ than the FWHM of each component.

\section{Discussion}
\label{sec:discuss}

\subsection{Reliability of Double \oiii}   
\label{subsec:reliability}

The ``good'' objects are definite cases of double-peaked \oiii\ profiles.  The split is confirmed
with good S/N and velocity agreement in $\lambda4959$, and in other
narrow lines when there is adequate S/N.  For the marginal cases, the confirmation is
less compelling in $\lambda4959$, and the S/N may be inadequate in other lines.  In order to
estimate the frequency with which noise might give the appearance of a double-peaked \oiii\ profile,
we examined a set of simulated single-peaked spectra for 1,773 SDSS quasars computed by \citet{salviander07} as a
test of their procedures.  These objects are typical of our double-peaked sample in terms
of \oiii\ strength and continuum S/N.  The simulated spectra reproduce the observed emission lines for each quasar, using a Gauss-Hermite fit to the lines in the original spectrum as described by \citet{salviander07}, along with a power-law continuum approximating the observed continuum
in the \oiii\ region.   Gaussian random noise was added pixel-by-pixel using the noise amplitude given in HDU3 of the SDSS spectrum.  For the present purposes, the simulated spectra were visually inspected in the same manner as our actual SDSS  parent sample.  In no case did a double-peaked $\lambda5007$ profile appear as a result of noise that would have met our standards even for  a ``marginal'' object.  We conclude that all of the ``good'' and most of the ``marginal'' objects are in fact genuine double-peaked \oiii\ AGN.

\subsection{Binary AGN?}
\label{subsec:binary}

Our search was motivated by the prospect of discovering binary AGN.  How many of our objects are actually binaries?    The Type 1 objects have broad emission lines and are clearly AGN.  The Type 2 objects have large equivalent widths of \oiii, a typical \oiii/\hbeta\ intensity ratio $\sim10$, characteristic of AGN, and sometimes show \heii~$\lambda4686$ or \nev.  This indicates that most are indeed AGN.  As noted above, alternative interpretations of double-peaked \oiii\ profiles include jet interactions and rotation.  One of our Type 2 objects, J1517+33, is spatially resolved but is clearly an example of a radio jet interaction with the NLR \citep{rosario10}.  Another of our objects, J1129+60, was observed with the VLA by Rosario et al.  It also has kpc scale radio structure and may involve jet interaction.  The majority of the radio undetected objects show double-peaked line profiles of a type that could reasonably occur for a binary AGN.  One AGN 1, J1307+46, shows two spatially resolved optical sources separated by 3~arcsec (15~kpc).   The others are unresolved in the SDSS images and may be candidates for binary AGN at spacings less than about 5~kpc, depending on redshift.  High resolution imaging in the optical, radio, and X-ray is needed to help determine which are actually binary AGN.    Pending such studies, can other arguments offer guidance as to which objects are binaries or disturbed NLRs, or at least indicate the proportions of such objects in our sample?

\subsubsection{Velocity separation}

The velocity separation of the double peaks is plausible for binary AGN in most cases.  For a circular orbit in the
gravitational potential of an isothermal sphere of velocity dispersion \sigstar, the orbital velocity is 
$2^{1/2}\,\sigstar$ \citep{binney88}.   For an eccentric orbit, the pericentric velocity could be several times \sigstar.  Our ``good''  radio-undetected Type 1 objects have a mean value $\mathrm{log}\,\mbh = 8.13$ solar masses, based on the broad line width and continuum luminosity using equation 2 of \citet{shields03}. (This refers to a subset of  13 objects with the best determined \mbh.)  This corresponds to a host bulge velocity dispersion $\mathrm{log}\,\sigstar = 2.30$ \kms\ based on the \mbhsigstar\ relationship as given by \citet{trem02}.  The mean  \oiii\ velocity split is $\mathrm{log}\,\Delta v = 2.52$ in \kms; $\Delta v/\sigstar$ averages 1.7 with a range of 1.0 to 6.  A precise prediction of $\Delta v/\sigstar$ for binary AGN with double \oiii\ is difficult because of unknown orbital parameters and a selection for binaries whose relative velocity vector is oriented parallel to the line of sight.  In any case,  the largest separations that we see, up to 1400~\kms, seem rather high for
mergers and may represent cases of a bipolar jet.

\subsubsection{Line intensity ratios}

The utility of emission line ratios to distinguish binaries from disturbed NLRs is unclear.
Jet interaction sources may involve gas dynamically affected by the jets but still photoionized by the AGN continuum and showing normal photoionization line ratios \citep[e.g.,][]{whittle05}.   Binarity could conceivably affect the NLR line ratios, particularly from spacings $<1$~kpc approaching the size of the NLR itself.   For our radio-undetected ``good'' Type 1 objects,  average values are $\mathrm{EW}(\lambda5007) = 72$~\AA\ and $F(\lambda3727)/F(\lambda5007) = 0.24$, where $\lambda3727$ refers to the combined intensity of the \oii\ doublet.  These values are not significantly different from a control sample of comparable non-double \oiii\ SDSS quasars, and the \oii/\oiii\ ratio is typical of power-law photoionized objects \citep{baldwin81}.  The \oi~$\lambda6300$ line is measureable in only a fraction of our objects.  For those objects, the average value of $I(\lambda6300)/I(\lambda5007)$ is 0.11,
similar to a comparable control sample; and these objects lie in the power-law photoionized region of the \oii/\oiii -- \oi/\oii  diagram of \citet{baldwin81}.  For our radio-detected ``good'' Type 1 objects,  average values are $\mathrm{EW}(\lambda5007) = 128$~\AA\ 
and $F(\lambda3727)/F(\lambda5007) = 0.17$.
The greater strength of the narrow emission lines in radio loud AGN is well known.  
The mean values for \oii/\oiii\ differ by less than $2~\sigma$ between the radio-detected and undetected samples.  
Given the impact of radio emission on the narrow lines for normal (non-double peaked) AGN, comparisons of line intensities 
for radio-detected and non-detected objects may not give clear guidance as to which double-peaked objects are binaries or jet-interactions.

Differential reddening of the red and blue components of the double-peaked narrow lines may offer a diagnostic
(M. Whittle, private communication). Such a reddening differential is expected for bipolar jets but not for binary AGN or a rotating disk geometry. If
there is distributed reddening in the NLR, the blueshifted
component, being on the near side of the AGN, may show less reddening than the redshifted component. 
Allowing for noise, blending, and redshift, only a handful of our
objects would allow a reliable assessment of the \halpha/\hbeta\ intensity ratio for the two components separately,
too few for a meaningful test.   For our objects with measured \oiii\ component fluxes, the red/blue ratio 
is $F_{\mathrm r}/F_{\mathrm b} = 0.90 \pm 0.10$ for the AGN~1 and $1.30 \pm 0.20$ for the AGN~2 (standard error of the mean).
These values give no clear indication of a systematic difference in extinction, but the uncertainty is large, reflecting the scatter among
individual objects in $F_{\mathrm r}/F_{\mathrm b}$.

In a study of narrow line AGN in SDSS with double-peaked \oiii\ profiles,  \citet{liu09} find that the intensity ratio of \oiii\ to \hbeta\ is typically rather similar in the red and blue velocity components.  This is suggestive of a common ionizing source and may favor jets or disks over binary AGN for most objects, at least in 
their sample.

\subsubsection{Virial test}

In an independent study of AGN 2 with double-peaked profiles, 
\citet{wang09} examine the relative velocity offset of the \oiii\ red and blue components  
($v_{\rm r}/v_{\rm b}$) in relation to their relative intensities ($F_{\rm r}/F_{\rm b})$.  
They find that the brighter component tends to be closer to the velocity of the host galaxy.  
They argue that this is consistent with orbital motion of a binary galactic nucleus, assuming that
the brighter emission-line component typically comes from the more massive black hole (and associated stellar cluster) 
that would normally have the lower orbital velocity, $F_{\rm r}/F_{\rm b} \propto (v_{\rm r}/v_{\rm b})^{-1}$.

\begin{figure}[ht]
\begin{center}
\plotone{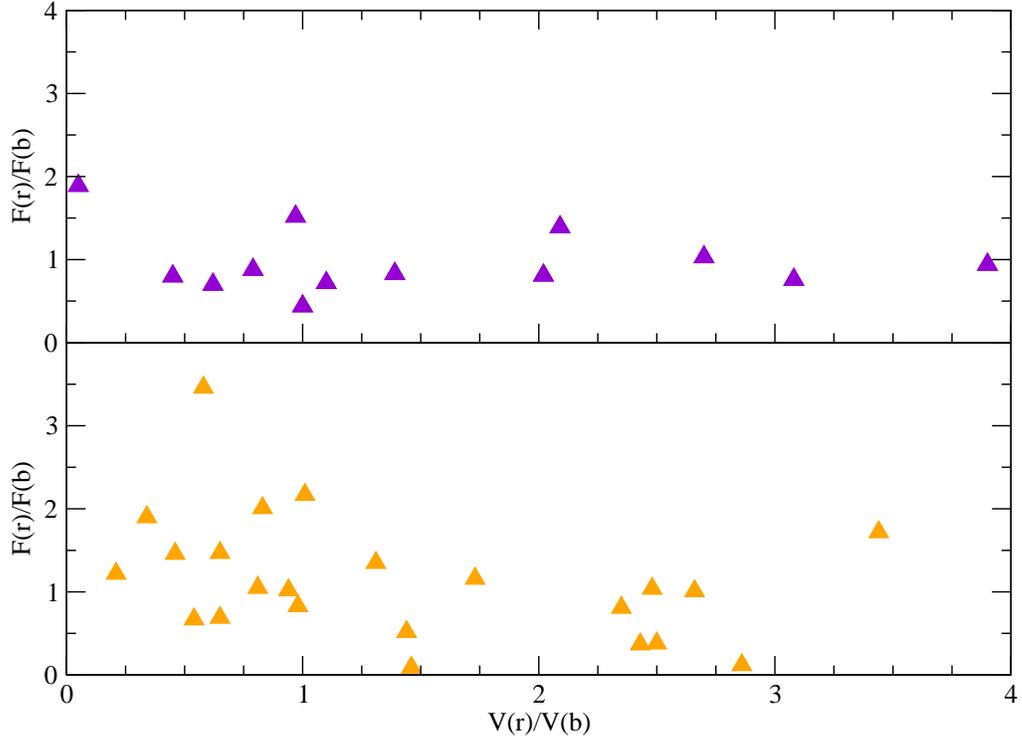}
\figcaption[]{
Flux ratio of the red and blue \oiii\ components as a function of the ratio of their velocity offsets relative to
the host galaxy redshift (see text).  Upper and lower panels show AGN 1 and AGN 2, respectively. 
\label{fig:virial} }
\end{center}
\end{figure}

We have examined our own data set for the trend found by Wang et al.  We considered separately the AGN 1 and AGN 2, using flux determinations for the two \oiii\ line components as given in Table 3.  The results are shown
in Figure \ref{fig:virial}.  There is little indication of the claimed virial trend for our AGN 1 sample.  Unfortunately, the number of useful AGN 1 is only 13.  This includes 7 objects with $v_{\rm r}/v_{\rm b} > 1$ and 6 objects with
$v_{\rm r}/v_{\rm b} < 1$.  For these AGN 1 and AGN 2, respectively, the average values are $v_{\rm r}/v_{\rm b} = (2.33, 0.65)$ and  $F_{\rm r}/F_{\rm b} = 0.92, 1.04$.   These values are consistent with no dependence of velocity offset on flux ratio at the $1~\sigma$ level, but they differ by $4~\sigma$ from the reciprocal relation predicted by the virial argument of Wang et al.   For our AGN 2, Figure \ref{fig:virial} suggests some inverse relationship between the velocity and flux ratios, considerably weaker than  $(v_{\rm r}/v_{\rm b})^{-1}$. However, the sample is small (23 useful objects), and our AGN~2 sample is subject to unknown biases, as discussed above.

\subsubsection{Luminosity dependence}

One of our objects, SDSS J131642.90+175332.5, was the subject of a detailed study by \citet{xu09}.  These authors consider alternative explanations of the double-peaked lines, including a binary AGN, biconical outflow, jet-cloud interaction, and other complexities in the NLR geometry. They suggest that binaries might be more common in quasars than in Seyfert galaxies, on the assumption that major mergers are more often involved in fueling quasars. We have examined our sample for such a trend. For luminosity bins ${\rm log}~\nu L_\nu(4400~{\rm \AA})$  in the ranges ($< $43.5, 43.5 - 44.0, 44.0 - 44.5, 44.5 - 45.0, $>$ 45.0), the number of AGN 1's in our sample is (4, 20, 34, 24, 4). These counts represent (0.35,  0.30,  0.39,  0.74, 1.13) percent of the number of objects in the parent sample in the corresponding luminosity bins.  For reference, the Seyfert/quasar boundary is 
$M_{\rm B} = -21.5 + 5~{\rm log}~h$  \citep{peterson97}, corresponding to   ${\rm log}~\nu L_\nu(4400~{\rm \AA}) =  44.44$.   The incidence of double peaks increases roughly a factor of two across our range of luminosities.   This resembles the factor of two increase in the incidence of double-peaked \oiii\ with increasing
\oiii\ luminosity found by \citet{liu09} in their study of AGN~2. This could be indicative of a substantial number of binaries among our sample. However, the origin of bipolar jets in AGN is not fully understood, and jets affecting the NLR could be more common in higher luminosity AGN. 

\subsubsection{AGN 1 versus AGN 2}

We suggest here another statistical argument involving the velocity splitting that may give an indication of the nature of the double-peaked narrow line AGN.  This involves a comparison of the typical velocity separation for Type 1 (broad lines) and Type 2 AGN (narrow lines only).  In the unified model of AGN \citep{urry95},  AGN 1 are viewed relatively close to the disk axis, whereas AGN 2  are viewed closer to the disk plane.  In AGN 2,  the ``dusty torus'' obscures the central continuum source and BLR.   Thus, the typical velocity projection onto the line of sight may differ between AGN~1 and AGN~2.  (1)  In the bipolar jet picture, the observed splitting is a function of the observer's location relative to the jet axis, $V = V_0 \cos{\theta}$ where $V_0$ is the true velocity separation.   In the unified model, on average, AGN 1 should have larger observed velocity separations than AGN 2.  Consider a simple model in which all objects have the same $V_0$, and in which all objects have a torus opening angle of 45 degrees.   Let $w_1, w_2$ represent  the average observed velocity splitting for AGN 1 and 2 respectively.   Then for AGN 1, $w_1 = V_0 \langle \cos{\theta} \rangle$, where the average is over polar angles of 0 to 45 degrees; and for AGN~2,  the range of $\theta$ is from 45 to 90 degrees.
This gives $\langle \cos{\theta} \rangle = 0.85$  for AGN 1 and 0.35 for AGN 2, so that $w_1/w_2 = 2.4$.   However, this exaggerates the effect, because there is a minimum splitting for the double-peaked profile to be resolved.  Our results suggest a value for this $V_{\mathrm {min}}$ of about  200~\kms.  For a rough estimate of this correction, let us take $V_0 = 500~\kms$, based on the mean splitting for our AGN 1.  Thus, AGN with $\cos{\theta} < 0.4$ will not show double peaks.  The range of polar angles for AGN 2 to show double peaks is then 45 to 66 degrees, giving  $\langle \cos{\theta} \rangle = 0.55$ and  $w_1/w_2 = 1.54  $.  
(2)  In a rotating disk model with circular velocity $V_{\mathrm {rot}}$, the observed velocity splitting is $V = V_0 \sin{\theta}$, where $V_0 = 2 V_{\mathrm {rot}}$.  By a similar procedure to that above, we find  $w_1/w_2 =0.53$ for $V_{\mathrm {min}} =  0$ or  $w_1/w_2 = 0.63$  using $V_{\mathrm {min}} = 200~\kms$.  
Thus, the sense of the difference between AGN 1 and 2 is reversed from the jet case, but the magnitude of the effect is similar.  
(3)  For binaries, the prediction is less clear.  Observationally, only one of the two AGN need be a Type 1 for broad lines to be observed.  If the dusty tori of both objects tend to be coplanar with the orbit, then the situation resembles case (2) above for a rotating disk, and AGN 1 might show smaller velocity separations than AGN 2.   If there is little correlation between the orbital plane and the dusty tori, then $w_1$ and $w_2$ should be similar.

 Our observed double-peaked \oiii\ objects have $w_1 =  434\pm25~\kms$ for AGN 1 and $480\pm 22~\kms$ for AGN 2 (standard error of the mean).  The AGN 1 and 2 samples have closely similar mean luminosities in \oiii, consistent with similar underlying AGN luminosities.  The observed ratio of $w_1/w_2 = 0.90\pm0.07$ disagrees with a predominance of either bipolar jets or disks, if the above estimates are roughly correct.  This result, by itself, might suggest either similar numbers of disks and jets, or a predominance of binaries.   We refrain from drawing conclusions, however, because of possible biases related to the presence of Type 2 AGN in our sample.

The double-peaked objects of \citet{liu09}, all AGN 2, have $w_2 = 368\pm8~\kms$, close to our value.   
Our sample has a mean redshift of $z = 0.37$ and 0.27 for AGN 1 and 2, respectively.  For Liu et al. the mean redshift is 0.16, and only 0.088 for Wang et al.  
For a subset of our lower redshift type 1 AGN having an average redshift of 0.158, similar to that of Liu et al., we find that our average velocity splitting is $w_1=425\pm58~\kms$, within $1~\sigma$ of Liu et al.'s sample, and that $w_1/w_2=1.1$, again consistent with binaries. If we compare Liu et al.'s AGN~2 to our full sample of Type 1 objects, we find $w_1/w_2=1.2$. 
 The objects of \citet{wang09} have $w_2 = 344~\kms$, but their sample has a much smaller average redshift of 0.088, and their mean \oiii\ luminosity $10^{41.30}~\ergps$  is an order-of-magnitude smaller than for our sample.

\subsection{Line-of-sight correction}
\label{subsec:line}

The width of the narrow emission lines correlates with \sigstar\ empirically,  with
$\mathrm{log}\,\sigthree/\sigstar \approx 0.0\pm0.15$, where 
$\sigthree \equiv \mathrm{FWHM}(\lambda5007)/2.35$  \citep{nelson96, bonning05}.  
The profile will appear double-peaked if the velocity separation is greater than $\sim \sigthree$.  Given that the orbital velocity of a binary should be of this order (see above), one might expect  that some binaries will  show a double-peaked \oiii\ profile and others will not.   Moreover,  a substantial fraction of binaries may be at a low velocity phase of an eccentric orbit, and many will have their relative velocity vector close to the plane of the sky.  The true number of binaries (or otherwise disturbed NLRs) may be several times larger than the number of double-peaked objects observed.  \citet{zhou04} estimate that this correction may be an order-of-magnitude.  
Thus, if a large fraction of our radio-undetected Type 1 objects are binaries, then the implied true incidence  of binaries is as large as $\sim10\%$.
However, such a large fraction of binaries appears to conflict with the statistics of spatially resolved AGN (see below).  This in turn suggests that only a small fraction of the double-opeaked \oiii\ objects are in fact binary AGN.

\subsection{Fueling of Binary AGN}
\label{subsec:fueling}

The role of mergers in fueling AGN activity is an important topic of study.  While seemingly undisturbed AGN are observed, high resolution imaging of quasars shows close companions and and features of gravitational interaction in many cases \citep[e.g.,][]{bahcall97,bennert08}.  Simulations of galactic collisions with gas show that tidal torques lead to concentrations of gas in each nucleus after the first close encounter, and in the nucleus of the final merger product \citep[and references therein]{hopkins06}.   It seems likely that a large fraction of luminous AGN are triggered by mergers.  This suggests that binary AGN would be a common occurrence.

As noted above, the incidence of known close binaries at 1 or 2~kpc separation is  only
$\sim10^{-3}$, and it has been argued that this is lower than expected if both black holes in a merger
are fueled independently \citep{junkkarinen01}.   A similar conclusion is also reached by \citet{foreman09}, in a study of optically resolved binaries in the 10 to 100~kpc range.  In our case, a typical redshift is $z \sim 0.3$, and a 1~arcsec limit on our unresolved objects corresponds to
$\sim 5$~kpc.    What might one expect for the number of binaries in this range of separation?
For an orbital velocity of $300~\kms$ and radius of 3~kpc, the orbital period is $\torb \approx 10^{7.8}$~yr.  Dynamical friction causes the orbit to decay on a timescle \tdf\ of a few times \torb; for
typical parameters for our objects, $\tdf \approx 10^{8.9}~\yr$ \citep{junkkarinen01}.  The lifetime of an AGN outburst is often estimated to be the Salpeter growth timescale $\tagn \approx 10^{7.6}$~yr.   Thus,  the duration of an AGN outburst could be a substantial fraction of $t_{\mathrm{df}}$.  The fact that the galaxy is observed as an AGN implies that at least one black hole is currently fueled.  If fueling of each hole occurs independently, then the probability that a given AGN will have both holes fueled as a binary AGN is $p_{bin} \approx \tagn / \tdf  \approx 10^{-1.3}$.   This assumes that most AGN are triggered by mergers.

This theoretical estimate is considerably larger than our observed $\sim1\%$ incidence of double \oiii\ objects but similar to the above $10\%$ estimate after correction for unresolved double \oiii\ objects.  It is doubtful that the true incidence of binary AGN is so high, however. The number of resolved binary QSOs at the 2~kpc  scale and the 30~kpc scale is $\sim10^{-3}$ in both cases, based on objects at typical redshifts of 1 to 3 \citep{junkkarinen01, foreman09}.
If these indications of a low binary rate carry over to our redshift and separation range, this may imply that most of the double \oiii\ objects are in fact single AGN with a disturbed NLR.  A low number of optically
visible binary AGN  could result from a delay in fueling of the black holes until the orbital separation
has decayed below the kpc scale, or obscuration of most binary AGN at kpc separations by massive amounts of dusty gas \citep{hopkins06}.  In this context, it is interesting that ultraluminous infrared galaxies (ULIRGs) often show nuclear spacings of one or several kpc \citep{sanders96}.

\section{Conclusion}
\label{sec:conclusion}

We have found that approximately 1\%\ of broad-line SDSS quasars at $0.1 < z < 0.7$ have double-peaked \oiii\ line profiles, and other narrow lines show similar velocity splittings when the data quality is sufficient.   In most cases, the magnitude of the velocities is consistent with orbital motion in binary AGN with distinct NLRs.   However, many of the objects, including those with FIRST radio detections, are likely disturbed NLRs involving bipolar jets or other complexities.   
We have discussed a number of potential statistical tests of which mechanism accounts for most of the double-peaked objects, but available
data are not definitive.

AGN with double-peaked narrow lines are interesting individual targets for detailed study.  In addition, they offer a potential way to assess the probability of 
simultaneous fueling of both black holes in a galactic merger.  In order to realize this potential, further study of these objects, including high resolution imaging in the radio, optical, and X-ray, is needed to determine which of them are actually binary AGN.

\acknowledgments

We thank Amanda Bauer, Sarah Busch, Pamela Jean, Anandi Salinas, and Dan Vandenberk for assistance with an earlier version of this project.  
G.S. gratefully acknowledges support from the Jane and Roland Blumberg Centennial Professorship in Astronomy at the University of Texas at Austin.

Funding for the Sloan Digital Sky Survey (SDSS) has been provided by the Alfred P. Sloan Foundation, the Participating Institutions, the National Aeronautics and Space Administration, the National Science Foundation, the U.S. Department of Energy, the Japanese Monbukagakusho, and the Max Planck Society. The SDSS Web site is http://www.sdss.org/. The SDSS is managed by the Astrophysical Research Consortium (ARC) for the Participating Institutions. The Participating Institutions are The University of Chicago, Fermilab, the Institute for Advanced Study, the Japan Participation Group, The Johns Hopkins University, the Korean Scientist Group, Los Alamos National Laboratory, the Max-Planck-Institute for Astronomy (MPIA), the Max-Planck-Institute for Astrophysics (MPA), New Mexico State University, University of Pittsburgh, University of Portsmouth, Princeton University, the United States Naval Observatory, and the University of Washington.

\clearpage


\newpage

\newpage


\clearpage


\clearpage


\newpage


\begin{deluxetable}{lccccccc}
\tabletypesize{\footnotesize}
\label{tab:list}
\tablewidth{0pt}
\tablecaption{Double-Peaked AGN\label{t:tab1}}
\tablehead{
\colhead{SDSS Name}          &
\colhead{$z_{\rm SDSS}$}       &
\colhead{$v$} &
\colhead{log $\nu L_{\nu}$} &
\colhead{Spectral}  &
\colhead{Quality} &  
\colhead{$z_{\rm r}$} &
\colhead{$z_{\rm host}$}  \\
\colhead{ }            &
\colhead{ }            &
\colhead{(km/s)}            &
\colhead{(erg/s)}  &
\colhead{Type }  &
\colhead{ } &
\colhead{ } &
\colhead{ }}
\startdata

J011802.94-082647.2	&	0.137	&	350	&	43.69	&	2	&	m	&	0.13771	&	0.13713	\\
J012613.31+142013.4 	&	0.573	&	580	&	44.35	&	2	&	g	&	0.57338	&		\\
J020011.52-093126.1 	&	0.361	&	200	&	44.49	&	1	&	m	&	0.36077	&		\\
J072554.42+374436.9 	&	0.634	&	250	&	44.81	&	1	&	m	&	0.63500	&		\\
J074129.66+392835.9 	&	0.210	&	480	&	43.63	&	2	&	m	&	0.21059	&	0.20982	\\
J080315.67+483603.1	&	0.635	&	280	&	44.86	&	1	&	m	&	0.63576	&		\\
J080841.21+481351.9 	&	0.123	&	340	&	43.38	&	2	&	m	&	0.12388	&	0.12359	\\
J081507.41+430427.0 	&	0.510	&	500	&	44.09	&	1	&	g	&	0.51014	&		\\
J081542.53+063522.9 	&	0.244	&	280	&	44.12	&	1	&	m	&	0.24428	&		\\
J082357.80+391630.9 	&	0.166	&	700	&	43.67	&	1	&	m	&	0.16774	&	0.16630	\\
J082857.99+074255.7 	&	0.554	&	300	&	44.07	&	2	&	g	&	0.55475	&		\\
J084049.46+272704.7 	&	0.136	&	580	&	43.32	&	2	&	m	&	0.13706	&	0.13559	\\
J090246.93+012028.2 	&	0.513	&	240	&	43.88	&	2	&	m	&	0.51408	&		\\
J090615.92+121845.6	&	0.644	&	200	&	45.23	&	1	&	g	&	0.64410	&		\\
J090947.85+312443.6	&	0.264	&	1050	&	44.41	&	1	&	g	&	0.26449	&		\\
J090958.32+085542.2 	&	0.158	&	240	&	43.20	&	2	&	m	&	0.15830	&	0.15869	\\
J091110.19+140632.6	&	0.387	&	350	&	44.87	&	1	&	g	&	0.38744	&		\\
J091459.25+210219.6	&	0.133	&	540	&	43.32	&	2	&	g	&	0.13418	&	0.13335	\\
J091649.41+000031.5 	&	0.222	&	310	&	43.47	&	1	&	m	&	0.22307	&	0.22223	\\
J091654.09+521723.0 	&	0.219	&	315	&	43.70	&	1	&	m	&	0.21998	&		\\
J092152.46+515348.0 	&	0.587	&	440	&	44.06	&	2	&	m	&	0.58806	&		\\
J092455.24+051052.0	&	0.149	&	410	&	43.85	&	2	&	m	&	0.15005	&	0.15042	\\
J094100.81+143614.4	&	0.383	&	870	&	44.20	&	2	&	m	&	0.38493	&		\\
J094144.83+575123.6	&	0.159	&	300	&	43.69	&	1	&	g	&	0.15924	&	0.15866	\\
J094205.83+125433.6*	&	0.154	&	680	&	43.95	&	2	&	g	&	0.15443	&	0.15420	\\
J094236.68+192541.1	&	0.540	&	480	&	45.23	&	1	&	g	&	0.54249	&		\\
J095207.62+255257.2	&	0.339	&	475	&	44.35	&	1	&	g	&	0.33986	&	0.33892	\\
J100145.30+283330.3*	&	0.114	&	550	&	43.15	&	2	&	g	&	0.11543	&	0.11550	\\
J100708.01+242039.0	&	0.544	&	250	&	44.62	&	1	&	g	&	0.54497	&		\\
J101034.28+372514.7 	&	0.282	&	1400	&	43.98	&	1	&	g	&	0.28228	&		\\
J101241.20+215556.0	&	0.111	&	210	&	43.51	&	1	&	m	&	0.11126	&	0.11069	\\
J102004.36+324342.5	&	0.484	&	340	&	44.26	&	1	&	g	&	0.48484	&		\\
J102038.74+212806.5	&	0.137	&	660	&	42.53	&	2	&	m	&	0.13925	&		\\
J102045.58+030306.9 	&	0.535	&	260	&	44.54	&	1	&	m	&	0.53540	&		\\
J102727.90+305902.4	&	0.124	&	360	&	43.81	&	2	&	g	&	0.12488	&	0.12451	\\
J103138.67+380651.7	&	0.492	&	730	&	44.82	&	1	&	g	&	0.49211	&		\\
J103326.92+284751.0	&	0.591	&	390	&	44.66	&	2	&	g	&	0.59217	&		\\
J103752.23+312500.3 	&	0.160	&	900	&	43.66	&	2	&	m	&	0.16275	&	0.16020	\\
J104257.96+385347.2	&	0.107	&	440	&	43.38	&	2	&	g	&	0.10716	&	0.10661	\\
J105104.54+625159.3	&	0.436	&	360	&	44.99	&	1	&	m	&	0.43734	&		\\
J111013.20+053338.8	&	0.152	&	830	&	43.46	&	2	&	m	&	0.15271	&		\\
J111710.07+333950.3	&	0.128	&	560	&	43.69	&	2	&	m	&	0.12809	&	0.12786	\\
J112019.61+130320.0 	&	0.314	&	600	&	44.13	&	1	&	m	&	0.31486	&		\\
J112319.21+302825.4	&	0.522	&	480	&	43.50	&	2	&	g	&	0.52294	&		\\
J112507.33+023719.1 	&	0.260	&	400	&	43.92	&	2	&	m	&	0.26095	&	0.25989	\\
J112634.84+455935.7 	&	0.278	&	300	&	43.86	&	2	&	m	&	0.27884	&	0.27849	\\
J112939.78+605742.6 	&	0.112	&	550	&	43.61	&	2	&	g	&	0.11276	&	0.11171	\\
J113020.99+022211.5 	&	0.241	&	330	&	44.19	&	1	&	m	&	0.24194	&		\\
J113045.33+505509.1 	&	0.592	&	270	&	44.69	&	1	&	m	&	0.59207	&		\\
J113105.07+610405.1 	&	0.338	&	330	&	44.48	&	1	&	m	&	0.33767	&		\\
J113257.84+604653.6 	&	0.233	&	330	&	43.44	&	1	&	g	&	0.23330	&		\\
J114852.65+151415.8 	&	0.113	&	260	&	43.49	&	1	&	m	&	0.11405	&	0.11325	\\
J114908.49+144547.0 	&	0.595	&	200	&	44.67	&	1	&	m	&	0.59668	&		\\
J115106.69+471157.7 	&	0.318	&	490	&	44.62	&	1	&	m	&	0.31794	&		\\
J115523.74+150756.9 	&	0.287	&	560	&	44.40	&	1	&	g	&	0.28769	&		\\
J115713.07+515511.5 	&	0.564	&	220	&	44.78	&	2	&	g	&	0.56505	&		\\
J115714.97+081632.0*	&	0.201	&	420	&	43.70	&	1	&	g	&	0.20205	&	0.20122	\\
J120240.68+263138.6	&	0.476	&	450	&	45.21	&	1	&	g	&	0.47806	&		\\
J120343.22+283557.8	&	0.374	&	850	&	44.21	&	1	&	g	&	0.37662	&		\\
J120526.04+321314.6 	&	0.485	&	670	&	43.75	&	2	&	g	&	0.48684	&		\\
J120704.51+384024.7 	&	0.572	&	390	&	44.81	&	1	&	g	&	0.57316	&		\\
J120725.59+460205.1 	&	0.213	&	490	&	43.60	&	2	&	m	&	0.21485	&	0.21331	\\
J121607.34-021417.7 	&	0.100	&	300	&	43.64	&	1	&	m	&	0.10117	&		\\
J121659.94+323106.0	&	0.125	&	210	&	43.79	&	1	&	m	&	0.12595	&	0.12554	\\
J121756.47+380022.7 	&	0.214	&	680	&	43.86	&	1	&	m	&	0.21485	&	0.21466	\\
J121911.16+042905.9 	&	0.555	&	480	&	44.75	&	1	&	g	&	0.55757	&		\\
J122709.83+124854.5 	&	0.194	&	670	&	43.60	&	2	&	g	&	0.19542	&	0.19389	\\
J123605.45-014119.1 	&	0.211	&	330	&	43.65	&	2	&	g	&	0.21201	&	0.21110	\\
J123915.40+531414.6 	&	0.201	&	800	&	43.96	&	1	&	g	&	0.20359	&	0.20200	\\
J124046.63+512902.1 	&	0.597	&	260	&	44.12	&	1	&	g	&	0.59816	&		\\
J124504.19+372300.1* 	&	0.279	&	550	&	43.52	&	2	&	g	&	0.27966	&	0.28052	\\
J124813.82+362423.6	&	0.207	&	350	&	43.94	&	1	&	g	&	0.20772	&	0.20690	\\
J124859.72-025730.7 	&	0.487	&	345	&	44.60	&	1	&	g	&	0.48772	&		\\
J124928.36+353926.8	&	0.527	&	380	&	44.00	&	2	&	g	&	0.52923	&		\\
J125327.50+254747.4	&	0.483	&	680	&	44.26	&	1	&	m	&	0.48532	&		\\
J125439.39+021100.6 	&	0.404	&	225	&	44.10	&	1	&	m	&	0.40492	&		\\
J130724.08+460400.9*	&	0.353	&	580	&	43.98	&	1	&	g	&	0.35382	&	0.35174	\\
J131018.47+250329.5	&	0.313	&	180	&	44.10	&	1.8	&	m	&	0.31339	&	0.31336	\\
J131611.76+310500.2	&	0.377	&	240	&	44.36	&	1	&	m	&	0.37720	&	0.37719	\\
J131642.90+175332.5*	&	0.150	&	450	&	43.77	&	2	&	g	&	0.15093	&	0.14999	\\
J132318.82+030807.1	&	0.269	&	615	&	43.91	&	2	&	g	&	0.26990	&	0.26871	\\
J132701.41+202306.1	&	0.571	&	350	&	44.36	&	2	&	m	&	0.57118	&		\\
J132855.78+213532.5	&	0.135	&	550	&	43.70	&	2	&	g	&	0.13580	&	0.13468	\\
J133226.34+060627.3	&	0.207	&	420	&	43.87	&	2	&	g	&	0.20738	&	0.20653	\\
J133455.24+612042.1	&	0.495	&	215	&	44.31	&	1	&	g	&	0.49561	&		\\
J134415.75+331719.1	&	0.686	&	730	&	44.81	&	2	&	g	&	0.68619	&		\\
J135024.66+240251.4	&	0.557	&	240	&	44.49	&	1	&	m	&	0.55816	&		\\
J140209.36+621025.8	&	0.330	&	280	&	44.15	&	1	&	g	&	0.33052	&		\\
J140318.10+164959.6	&	0.455	&	340	&	44.46	&	1	&	m	&	0.45653	&		\\
J140500.14+073014.1	&	0.135	&	340	&	43.39	&	2	&	m	&	0.13572	&	0.13506	\\
J140646.11+234821.0	&	0.519	&	310	&	44.55	&	1	&	m	&	0.51943	&		\\
J140816.02+015528.3 	&	0.166	&	350	&	43.39	&	2	&	g	&	0.16649	&	0.16587	\\
J140923.51-012430.5	&	0.405	&	270	&	44.21	&	1	&	g	&	0.40548	&		\\
J141316.06+020346.9 	&	0.507	&	540	&	44.43	&	1	&	m	&	0.50889	&		\\
J141431.04-003042.9 	&	0.138	&	400	&	43.41	&	2	&	m	&	0.13893	&	0.13783	\\
J141445.99+370202.1	&	0.260	&	340	&	44.09	&	1	&	g	&	0.26069	&		\\
J143135.43-011159.8 	&	0.560	&	490	&	44.55	&	1	&	g	&	0.56214	&		\\
J144012.74+615633.0	&	0.275	&	800	&	44.57	&	1	&	g	&	0.27619	&		\\
J144102.38+390114.5	&	0.176	&	290	&	43.80	&	2	&	m	&	0.17685	&	0.17614	\\
J144105.64+180507.9	&	0.107	&	280	&	43.59	&	1	&	m	&	0.10736	&	0.10668	\\
J144157.24+094859.1* 	&	0.220	&	810	&	43.76	&	2	&	g	&	0.22238	&	0.21986	\\
J144541.30+334107.8 	&	0.131	&	400	&	43.52	&	2	&	g	&	0.13156	&	0.13100	\\
J144748.79+624444.7 	&	0.230	&	870	&	43.87	&	1	&	m	&	0.23036	&		\\
J145110.04+490813.5	&	0.156	&	220	&	43.81	&	1	&	m	&	0.15632	&	0.15606	\\
J145336.31+204357.5	&	0.116	&	520	&	43.58	&	2	&	g	&	0.11699	&	0.11572	\\
J145408.36+240521.3	&	0.535	&	300	&	44.64	&	1	&	g	&	0.53578	&		\\
J145538.76+401913.0 	&	0.461	&	240	&	44.38	&	1	&	g	&	0.46170	&		\\
J145717.69+110412.4	&	0.462	&	200	&	44.21	&	1	&	m	&	0.46210	&		\\
J150125.57+111356.6	&	0.151	&	940	&	43.89	&	1	&	m	&	0.15179	&	0.15179	\\
J150243.09+111557.3	&	0.390	&	720	&	44.51	&	1	&	g	&	0.39326	&		\\
J150437.67+541149.6 	&	0.305	&	700	&	44.43	&	1	&	g	&	0.30770	&		\\
J151518.29+551535.3 	&	0.513	&	250	&	44.93	&	1	&	m	&	0.51436	&		\\
J151656.59+183021.5	&	0.580	&	360	&	44.28	&	2	&	g	&	0.58167	&		\\
J151709.20+335324.7* 	&	0.135	&	700	&	43.39	&	1	&	g	&	0.13664	&	0.13562	\\
J151735.17+214532.5	&	0.399	&	500	&	44.45	&	1	&	g	&	0.40092	&		\\
J151757.36+114452.6	&	0.227	&	400	&	43.79	&	2	&	g	&	0.22767	&	0.22701	\\
J151842.95+244026.0	&	0.561	&	280	&	44.43	&	2	&	g	&	0.56174	&	0.55550	\\
J151944.90+191353.3	&	0.245	&	600	&	43.87	&	1	&	m 	&	0.24500	&	0.24488	\\
J152117.30+075955.4	&	0.463	&	500	&	44.78	&	1	&	m 	&	0.46382	&		\\
J152327.57+262940.7	&	0.236	&	540	&	44.05	&	2	&	m	&	0.23723	&	0.23636	\\
J152431.41+323750.6 	&	0.629	&	350	&	43.81	&	2	&	g	&	0.63083	&		\\
J152506.63+022425.1	&	0.337	&	320	&	44.79	&	1	&	m	&	0.33783	&		\\
J153231.80+420342.7 	&	0.210	&	300	&	44.02	&	1	&	g	&	0.21002	&	0.20948	\\
J153301.43+070513.7	&	0.354	&	510	&	44.30	&	1	&	m 	&	0.35340	&		\\
J153423.19+540809.0	&	0.215	&	290	&	43.68	&	1	&	g	&	0.21553	&	0.21464	\\
J153714.71+121150.8	&	0.148	&	320	&	43.77	&	2	&	m	&	0.14842	&	0.14770	\\
J153944.11+343503.9	&	0.551	&	400	&	44.18	&	2	&	g	&	0.55237	&		\\
J154040.50+185402.9	&	0.137	&	180	&	43.27	&	1.9	&	g	&	0.13761	&	0.13705	\\
J154107.81+203608.8	&	0.508	&	400	&	44.32	&	1	&	g	&	0.50924	&		\\
J154637.12+122832.5	&	0.386	&	430	&	44.30	&	1	&	m	&	0.39956	&		\\
J154713.92+103359.8	&	0.638	&	360	&	44.70	&	1	&	m	&	0.63887	&		\\
J155634.15+105616.7	&	0.449	&	420	&	44.48	&	1	&	m	&	0.44990	&		\\
J155645.97+241828.5	&	0.220	&	530	&	44.12	&	2	&	g	&	0.22078	&	0.21950	\\
J160659.57+083514.6 	&	0.187	&	380	&	43.76	&	2	&	m	&	0.18745	&	0.18661	\\
J161027.41+130806.8	&	0.229	&	290	&	43.97	&	1	&	g	&	0.22948	&	0.22868	\\
J161141.95+495847.9 	&	0.116	&	265	&	43.59	&	1	&	m	&	0.11689	&	0.11643	\\
J161826.93+081950.7 	&	0.446	&	340	&	45.05	&	1	&	g	&	0.44655	&		\\
J161847.93+215925.4	&	0.335	&	320	&	44.71	&	1	&	g	&	0.33573	&	0.33429	\\
J161925.50+161032.9	&	0.134	&	540	&	43.62	&	2	&	g	&	0.13444	&	0.13340	\\
J162345.20+080851.1	&	0.199	&	470	&	43.98	&	2	&	g	&	0.19941	&	0.19857	\\
J165603.68+261722.1 	&	0.467	&	210	&	44.16	&	1	&	m	&	0.46697	&		\\
J170056.01+243928.2 	&	0.508	&	960	&	44.99	&	1	&	m	&	0.50918	&		\\
J171544.02+600835.4	&	0.157	&	350	&	43.70	&	2	&	g	&	0.15728	&	0.15648	\\
J171647.40+310403.1 	&	0.275	&	345	&	43.89	&	1.9	&	m	&	0.27589	&	0.27471	\\
J171850.29+304201.5 	&	0.282	&	700	&	44.23	&	1	&	m	&	0.28226	&		\\
J171930.56+293412.8	&	0.180	&	1000	&	43.81	&	2	&	g	&	0.18432	&		\\
J172507.11+274038.4 	&	0.490	&	400	&	44.19	&	1	&	m	&	0.49115	&		\\
J210449.13-000919.1 	&	0.135	&	380	&	43.41	&	2	&	m	&	0.13596	&	0.13523	\\

 \enddata

\tablecomments{AGN with double-peaked narrow emission lines from the Sloan Digital Sky Survey, including redshift as given by SDSS, rest-frame velocity separation of the two peaks of \oiii~$\lambda5007$, optical continuum luminosity from the SDSS spectrum,  AGN type, quality of the double peak as described in Section 2, redshift of the red peak of $\lambda5007$, and host galaxy redshift (when measurable).   Objects with an asterisk following their name are accompanied by an image in Figure \ref{fig:images}.  See text for discussion.}

\end{deluxetable}


\begin{deluxetable}{lccccc}
\tablewidth{0pt}
\tablecaption{Radio Flux and Luminosity\label{t:tab2}}
\tablehead{
\colhead{SDSS Name}          &
\colhead{$z_{\rm SDSS}$}       &
\colhead{$S_{\nu}$} &
\colhead{log $L_{\nu}$}         &
\colhead{log $R$} & 
\colhead{Spectral Type}  \\
\colhead{ }            &
\colhead{ }            &
\colhead{(mJy)}            &
\colhead{(\ergshz)}      &
\colhead{ } &
\colhead{ }} 
\startdata

J080315.67+483603.1	&	0.635	&	2.73	&	31.08	&	1.05	&	1	\\
J081507.41+430427.0 	&	0.510	&	5.44	&	31.16	&	1.90	&	1	\\
J090947.85+312443.6	&	0.264	&	1.81	&	30.03	&	0.46	&	1	\\
J091110.19+140632.6	&	0.387	&	5.41	&	30.88	&	0.85	&	1	\\
J094236.68+192541.1	&	0.540	&	1.35	&	30.61	&	0.21	&	1	\\
J105104.54+625159.3	&	0.436	&	1.74	&	30.51	&	0.35	&	1	\\
J112019.61+130320.0 	&	0.314	&	1.64	&	30.16	&	0.87	&	1	\\
J115106.69+471157.7 	&	0.318	&	3.17	&	30.46	&	0.67	&	1	\\
J115523.74+150756.9 	&	0.287	&	1.64	&	30.07	&	0.51	&	1	\\
J120240.68+263138.6	&	0.476	&	61.67	&	32.14	&	1.77	&	1	\\
J124813.82+362423.6	&	0.207	&	2.36	&	29.92	&	0.81	&	1	\\
J125327.50+254747.4	&	0.483	&	2.87	&	30.83	&	1.40	&	1	\\
J144012.74+615633.0	&	0.275	&	3.46	&	30.36	&	0.62	&	1	\\
J145110.04+490813.5	&	0.156	&	1.17	&	29.35	&	0.37	&	1	\\
J150243.09+111557.3	&	0.390	&	9.2	&	31.12	&	1.45	&	1	\\
J151518.29+551535.3 	&	0.513	&	2.29	&	30.79	&	0.70	&	1	\\
J151709.20+335324.7 	&	0.135	&	106.7	&	31.17	&	2.62	&	1	\\
J152506.63+022425.1	&	0.337	&	2.72	&	30.45	&	0.49	&	1	\\
J154713.92+103359.8	&	0.638	&	49.84	&	32.35	&	2.48	&	1	\\
J161141.95+495847.9 	&	0.116	&	1.44	&	29.16	&	0.40	&	1	\\
J161826.93+081950.7 	&	0.446	&	135.23	&	32.42	&	2.20	&	1	\\
J161847.93+215925.4	&	0.335	&	2.76	&	30.45	&	0.57	&	1	\\
J171850.29+304201.5 	&	0.282	&	1.25	&	29.94	&	0.54	&	1	\\
J074129.66+392835.9 	&	0.210	&	2.84	&	30.01	&	1.22	&	2	\\
J090246.93+012028.2 	&	0.513	&	1.19	&	30.51	&	1.46	&	2	\\
J090958.32+085542.2 	&	0.158	&	2.66	&	29.72	&	1.35	&	2	\\
J091459.25+210219.6	&	0.133	&	2.76	&	29.57	&	1.08	&	2	\\
J092152.46+515348.0 	&	0.587	&	2.48	&	30.96	&	1.74	&	2	\\
J094100.81+143614.4	&	0.383	&	7.39	&	31.01	&	1.64	&	2	\\
J094205.83+125433.6	&	0.154	&	6.14	&	30.05	&	0.94	&	2	\\
J102038.74+212806.5	&	0.137	&	5.33	&	29.88	&	2.19	&	2	\\
J102727.90+305902.4	&	0.124	&	1.28	&	29.17	&	0.20	&	2	\\
J103326.92+284751.0	&	0.591	&	1.21	&	30.65	&	0.83	&	2	\\
J103752.23+312500.3 	&	0.160	&	1.7	&	29.53	&	0.71	&	2	\\
J111710.07+333950.3	&	0.128	&	1.55	&	29.29	&	0.43	&	2	\\
J112319.21+302825.4	&	0.522	&	1.15	&	30.51	&	1.84	&	2	\\
J112939.78+605742.6 	&	0.112	&	25.69	&	30.38	&	1.60	&	2	\\
J115713.07+515511.5 	&	0.564	&	4.8	&	31.21	&	1.26	&	2	\\
J120526.04+321314.6 	&	0.485	&	1.5	&	30.55	&	1.64	&	2	\\
J124928.36+353926.8	&	0.527	&	2.7	&	30.89	&	1.72	&	2	\\
J131642.90+175332.5	&	0.150	&	10.66	&	30.27	&	1.33	&	2	\\
J132855.78+213532.5	&	0.135	&	5.68	&	29.90	&	1.03	&	2	\\
J134415.75+331719.1	&	0.686	&	9.32	&	31.69	&	1.71	&	2	\\
J140500.14+073014.1	&	0.135	&	4.58	&	29.80	&	1.25	&	2	\\
J140816.02+015528.3 	&	0.166	&	1.1	&	29.38	&	0.82	&	2	\\
J141431.04-003042.9 	&	0.138	&	2.03	&	29.47	&	0.90	&	2	\\
J144157.24+094859.1 	&	0.220	&	6.03	&	30.38	&	1.46	&	2	\\
J144541.30+334107.8 	&	0.131	&	4.38	&	29.76	&	1.07	&	2	\\
J145336.31+204357.5	&	0.116	&	1.22	&	29.09	&	0.35	&	2	\\
J151656.59+183021.5	&	0.580	&	1296.39	&	33.67	&	4.22	&	2	\\
J162345.20+080851.1	&	0.199	&	1.95	&	29.80	&	0.65	&	2	\\
J171544.02+600835.4	&	0.157	&	13.22	&	30.41	&	1.54	&	2	\\

\enddata

\tablecomments{Radio properties of the double-peaked AGN that have detections in the FIRST survey.
Columns give SDSS redshift, observed 20~cm flux density from FIRST, source specific luminosity at 6~cm rest wavelength
(assuming $L_\nu \propto \nu^{-0.7}$), radio loudness parameter $R$ \citep{kellermann89}, and AGN type.
}

\end{deluxetable}


\begin{deluxetable}{lcc}
\label{tab:list}
\tablewidth{0pt}
\tablecaption{$\lambda5007$ Component Fluxes\label{t:tab3}}
\tablehead{
\colhead{SDSS Name}          &
\colhead{$F_{\rm r}$}       &
\colhead{$F_{\rm b}$} \\
\colhead{ }            &
\colhead{ }            &
\colhead{ }}
\startdata

J080315.67+483603.1	&	51	&	69	\\
J081542.53+063522.9	&	73	&	83	\\
J082857.99+074255.7	&	247	&	169	\\
J084049.46+272704.7	&	73	&	88	\\
J090246.93+012028.2	&	341	&	39	\\
J090615.92+121845.6	&	84	&	68	\\
J090947.85+312443.6	&	350	&	210	\\
J091459.25+210219.6	&	322	&	492	\\
J091649.41+000031.5	&	211	&	263	\\
J091654.09+521723.0	&	98	&	151	\\
J092455.24+051052.0	&	196	&	99	\\
J094100.81+143614.4	&	122	&	357	\\
J094144.83+575123.6	&	278	&	626	\\
J094205.83+125433.6	&	950	&	650	\\
J101241.20+215556.0	&	611	&	595	\\
J102004.36+324342.5	&	112	&	44	\\
J102038.74+212806.5	&	123	&	56	\\
J102045.58+030306.9	&	89	&	51	\\
J102727.90+305902.4	&	903	&	372	\\
J103326.92+284751.0	&	652	&	1157	\\
J103752.23+312500.3	&	78	&	372	\\
J104257.96+385347.2	&	261	&	181	\\
J111710.07+333950.3	&	160	&	56	\\
J112319.21+302825.4	&	506	&	331	\\
J112507.33+023719.1	&	264	&	77	\\
J112634.84+455935.7	&	122	&	49	\\
J113020.99+022211.5	&	49	&	395	\\
J115106.69+471157.7	&	505	&	2009	\\
J115714.97+081632.0	&	105	&	69	\\
J120343.22+283557.8	&	456	&	531	\\
J120526.04+321314.6	&	310	&	398	\\
J120725.59+460205.1	&	72	&	123	\\
J121659.94+323106.0	&	198	&	276	\\
J121911.16+042905.9	&	299	&	387	\\
J122709.83+124854.5	&	482	&	367	\\
J123605.45-014119.1	&	491	&	485	\\
J124046.63+512902.1	&	88	&	67	\\
J124504.19+372300.1	&	255	&	83	\\
J124813.82+362423.6	&	215	&	259	\\
J124859.72-025730.7	&	354	&	120	\\
J125327.50+254747.4	&	52	&	181	\\
J130724.08+460400.9	&	96	&	102	\\
J131018.47+250329.5	&	426	&	590	\\
J132855.78+213532.5	&	384	&	222	\\
J133226.34+060627.3	&	218	&	82	\\
J133455.24+612042.1	&	69	&	71	\\
J134415.75+331719.1	&	381	&	366	\\
J135024.66+240251.4	&	68	&	98	\\
J140209.36+621025.8	&	12	&	90	\\
J140500.14+073014.1	&	30	&	37	\\
J140816.02+015528.3	&	100	&	102	\\
J140923.51-012430.5	&	141	&	230	\\
J141445.99+370202.1	&	43	&	175	\\
J143135.43-011159.8	&	61	&	96	\\
J145110.04+490813.5	&	150	&	187	\\
J145336.31+204357.5	&	89	&	262	\\
J145408.36+240521.3	&	99	&	132	\\
J150437.67+541149.6	&	186	&	658	\\
J151518.29+551535.3	&	115	&	94	\\
J151709.20+335324.7	&	1714	&	2434	\\
J151757.36+114452.6	&	151	&	279	\\
J151842.95+244026.0	&	313	&	310	\\
J151944.90+191353.3	&	184	&	97	\\
J153231.80+420342.7	&	432	&	493	\\
J153423.19+540809.0	&	257	&	338	\\
J153944.11+343503.9	&	323	&	537	\\
J154107.81+203608.8	&	78	&	154	\\
J154713.92+103359.8	&	42	&	102	\\
J155634.15+105616.7	&	143	&	256	\\
J155645.97+241828.5	&	103	&	223	\\
J161027.41+130806.8	&	228	&	164	\\
J161847.93+215925.4	&	148	&	343	\\
J161925.50+161032.9	&	527	&	212	\\
J162345.20+080851.1	&	437	&	186	\\
J170056.01+243928.2	&	171	&	91	\\
J171544.02+600835.4	&	797	&	1218	\\
J172507.11+274038.4	&	42	&	205	\\
J210449.13-000919.1	&	101	&	107	\\
 
\enddata

\tablecomments{Fluxes of the red and blue components of the $\lambda5007$ emission line. Units of flux are $10^{-17}~{\rm erg~cm^{-2}~s^{-1}}$.}

\end{deluxetable}


\begin{thebibliography}{38}
\expandafter\ifx\csname natexlab\endcsname\relax\def\natexlab#1{#1}\fi

\bibitem[Bahcall et al.(1997)]{bahcall97} Bahcall, J. N., Kirhakos, S.,  Saxe, D. H., \& Schneider, D. P. 1997 \apj, 479, 642

\bibitem[Baldwin et al.(1981)]{baldwin81}
Baldwin, J. A., Phillips, M. M., \& Terlevich, R. 1981, \pasp, 93, 5

\bibitem[Becker et al.(1995)]{becker95}
Becker, R.H., White, R.L., \& Helfand, D.J. 1995, \apj, 450, 559

\bibitem[{{Begelman} {et~al.}(1980){Begelman}, {Blandford}, \&
  {Rees}}]{begelman80}
{Begelman}, M.~C., {Blandford}, R.~D., \& {Rees}, M.~J. 1980, \nat, 287, 307

 \bibitem[Bennert et al.(2008)]{bennert08}
   Bennert, N., et al. 2008, \apj, 677, 846
   
\bibitem[Binney \& Tremaine(1988)]{binney88}
 Binney, J., \& Tremaine, S. 1988, {\em Galactic Dynamics}, Princeton Univ. Press
 
\bibitem[Bonning et al.(2005)]{bonning05}
   Bonning, E. W., Shields, G. A., Salviander, S., \& McLure, R. J. 2005, \apj, 626, 89

\bibitem[Comerford et al.(2009a)]{comerford09a} Comerford, J.M. \etal\ 2009a, \apj, 698, 956

\bibitem[Comerford et al.(2009b)]{comerford09b} Comerford, J.M. \etal\ 2009b, \apjl, 702, L82

\bibitem[Foreman et al.(2009)]{foreman09} Foreman, G., Volonteri, M., \& Dotti, M.  2009, \apj, 693, 1554

\bibitem[Gerke et al.(2007)]{gerke07} Gerke, B. F. \etal\ 2000, \apjl, 660, L23

\bibitem[Green et al.(2010)]{green10} Green, P., et al. 2010, arXiv:1001.1738

\bibitem[Hopkins et al.(2006)]{hopkins06} Hopkins, P., \etal\ 2006, \apjs, 163, 1

\bibitem[Junkkarinen et al.(2001)]{junkkarinen01} Junkkarinen, V.  \etal\ 2001, \apj, 549, L155

\bibitem[Kellermann et al.(1989)]{kellermann89} Kellermann, K. I.,  Sramek, R., Schmidt, M. Shaffer, D. B., \& Green, R. F. 1989, \aj,  98,1195

\bibitem[Komossa(2003)]{komossa03} Komossa, S., et al.  2003,  \apjl, 582, L15

\bibitem[Komossa(2006)]{komossa06} Komossa, S. 2006, Mem. S. A. It., 77, 733

\bibitem[Liu et al.(2009)]{liu09} Liu, X., Shen, Y., Strauss, M. A., \& Greene, J. E. 2010, 708, L427

\bibitem[Nelson \& Whittle(1996)]{nelson96} Nelson, C. H., \& Whittle, M. 1996, \apj, 465, 96

\bibitem[Peterson(1997)]{peterson97} Peterson, B. 1997, An Introduction to Active Galactic Nuclei (Cambridge: Cambridge Univ. Press)

\bibitem[Rodriguez et~al.(2009)] {rodriguez09} Rodriguez, C., Taylor, G. B., Zavala, R. T., 
  Philstro\"om, Y. M., \& Peck, A. N. 2009, apj, 697, 37

\bibitem[Rosario et al.(2010)] {rosario10} Rosario, D., et al. 2010, submitted

\bibitem[{{Salviander} {et~al.}(2007){Salviander}, {Shields}, {Gebhardt}, \&
  {Bonning}}]{salviander07}
{Salviander}, S., {Shields}, G.~A., {Gebhardt}, K., \& {Bonning}, E.~W. 2007,
  \apj, 662, 131
  
    
\bibitem[Sanders \& Mirabel (1996)]{sanders96} Sanders, D. B., \&  Mirabel, I. F. 1996, \araa, 34, 749
  
\bibitem[Schneider et al.(2007)]{schneider07} Schneider, D. P. \etal\ 2007, \aj, 134, 102

\bibitem[Shields et al.(2003)] {shields03} Shields, G. A., et al. 2003, \apj, 583, 124

\bibitem[Stockton et al.(2007)]{stockton07} Stockton, A., Canalio, G., Fu, Hai, \& Keel, W.  
    2007, \apj, 659, 195

\bibitem[Tremaine et al.(2002)]{trem02} Tremaine, S., et al.
2002, \apj, 574, 740

\bibitem[Urry \& Padovani(1995)]{urry95} Urry, M., \& Padovani, P. 1995, \pasp, 107, 803

\bibitem[Wang et al.(2009)]{wang09} Wang, J., Chen, Y., Hu, C., Mao, W., Zhang, S., Bian, W. 2009, \apjl, 705, L76

\bibitem[Whittle et al.(2005)]{whittle05} Whittle, M., Rosario, D. J., Silverman, J. D., Nelson, C. H.,
  \& Wilson, A. S.  2005, \aj, 129, 104

\bibitem[Whittle \& Wilson(2004)]{whittle04} Whittle, M., \& Wilson, A. S.  2004, \aj, 127, 606

\bibitem[Xu \& Komossa(2009)]{xu09} Xu, D., \& Komossa, S. 2009, \apjl, L20-L24

\bibitem[Zhou et al.(2004)]{zhou04} Zhou, H., Wang, T., Zhang, X., Dong, X., \& Li, C. 
2004, \apjl, 604, L33


\end{thebibliography}
\end{document}